\documentclass{aa}  

\usepackage{graphicx}
\usepackage{natbib}
\bibpunct{(}{)}{;}{a}{}{,}
\usepackage{txfonts}
\begin{document}

   \titlerunning{Doppler images and the underlying dynamo}
   \authorrunning{J\"arvinen et al.}

   \title{Doppler images and the underlying dynamo\thanks{Partially based on observations made with the Nordic Optical Telescope, operated by the Nordic Optical Telescope Scientific Association at the Observatorio del Roque de los Muchachos, La Palma, Spain, of the Instituto de Astrof\'{\i}sica de Canarias.}\fnmsep\thanks{Based partly on STELLA SES data.}}
   \subtitle{The case of AF~Leporis}

   \author{S.\ P. J\"arvinen\inst{1}
          \and
          R. Arlt\inst{1}
          \and
          T. Hackman\inst{2,3}
          \and
          S.C. Marsden\inst{4}
          \and
          M. K\"uker\inst{1}
          \and
          I.V. Ilyin\inst{1}
          \and
          S.V. Berdyugina\inst{5}
          \and
          K.G. Strassmeier\inst{1}
          \and
          I.A. Waite\inst{4}
          }

   \institute{Leibniz-Institut f\"ur Astrophysik Potsdam, 
		An der Sternwarte 16, D-14482 Potsdam\\
              \email{sjarvinen@aip.de}
         \and
             Department of Physics, University of Helsinki, 
             PO Box 64, 00014, Finland
         \and
             Finnish Centre for Astronomy with ESO (FINCA), 
             University of Turku, V\"ais\"al\"antie 20, 21500 Piikki\"o, Finland
         \and
             Computational Engineering and Science Research Centre, 
             University of Southern Queensland, Toowoomba, 4350, Australia
         \and
             Kiepenheuer-Institut f\"ur Sonnenphysik,
	        Sch\"oneckstr. 6, D-79104 Freiburg
             }

   \date{Received 2014; accepted 2014}

  \abstract
  { The (Zeeman-)Doppler imaging studies of solar-type stars very often reveal large high-latitude spots. This also includes F stars that possess relatively shallow convection zones, indicating that the dynamo operating in these stars differs from the solar dynamo.}
  { We aim to determine whether mean-field dynamo models of late-F type dwarf stars can reproduce the surface features recovered in Doppler maps. In particular, we wish to test whether the models can reproduce the high-latitude spots observed on some F dwarfs.}
   { The photometric inversions and the surface temperature maps of AF~Lep were obtained using the Occamian-approach inversion technique. Low signal-to-noise spectroscopic data were improved by applying the least-squares deconvolution method. The locations of strong magnetic flux in the stellar tachocline as well as the surface fields obtained from mean-field dynamo solutions were compared with the observed surface temperature maps.
}
   {The photometric record of AF~Lep reveals both long- and short-term variability. However, the current data set is too short for cycle-length estimates. From the photometry, we have determined the rotation period of the star to be $0.9660 \pm 0.0023$ days. The surface temperature maps show a dominant, but evolving, high-latitude (around +65\degr) spot. Detailed study of the photometry reveals that sometimes the spot coverage varies only marginally over a long time, and at other times it varies rapidly. Of a suite of dynamo models, the model with a radiative interior rotating as fast as the convection zone at the equator delivered the highest compatibility with the obtained Doppler images.
}
   {}

   \maketitle

%

\section{Introduction}

Solar-type stars form a very broad class of stars from late-F to early-K type dwarfs and sub-giants, all containing a convective envelope over a radiative interior. Observations of these stars provide excellent constraints for theoretical dynamo models. Our understanding of the operation of the magnetic dynamo in such stars is based on the solar case, where the solar activity cycle is believed to be generated through dynamo action operating either in the convection zone or in the stably stratified layer beneath it (for a short overview of solar dynamo models see, e.g., \citealt{2014A&A...563A..18P}). One would expect stars with an internal structure similar to that of the Sun, that is, stars with convective envelopes, to show the same type of dynamo operation. However, (Zeeman-)Doppler imaging studies frequently find large spots at high latitudes and large regions of a near-surface azimuthal field (not seen on the solar surface). For very active stars, it has been suggested that instead of a solar-like dynamo, which probably works in the overshoot zone, a distributed $\alpha^{2}\Omega$- or $\alpha^{2}$-dynamo is likely to be present \citep{1989A&A...213..411B, 1995A&A...294..155M}.

F stars possess relatively shallow convection zones. Nevertheless, observations show via the Doppler imaging method that at least some of these stars have high-latitude and even polar spots (e.g., \object{AF~Lep}: \citealt{2006ASPC..358..401M}; \object{$\tau$~Boo}: \citealt{2009MNRAS.398.1383F}; one of the components in \object{$\sigma^{2}$}~CrB: \citealt{2003A&A...399..315S}). Furthermore, surface differential-rotation estimates imply that $\delta\Omega$ increases with decreasing depth of the convection zone (e.g., \citealt{2011MNRAS.413.1939M}; \citealt{2005MNRAS.357L...1B}). Theoretical models support this result (e.g., \citealt{2007AN....328.1050K}). 

AF~Lep (HR~1817, HD~35850) is an active, young, rapidly rotating, single solar-like star. The first photometric observations of the target are from the late 60s \citep{1972MNRAS.159..165S}, and it was also detected with \emph{EXOSAT} \citep{1991A&AS...87..233C}. It has been identified as a member of the $\beta$~Pictoris moving group, which has been estimated to have an age of $\sim$20~Myr (\citealt{2008A&A...480..735F} and references therein; \citealt{2014MNRAS.438L..11B}; \citealt{2014MNRAS.445.2169M}). The colours suggest a spectral classification of F8/9 \citep{1996A&AS..115...41C}. The high Li abundance ($\log N$(Li)=3.2), solar metallicity, and high rotation velocity ($v \sin i=50$~km~s$^{-1}$) are consistent with the star being a young object. Furthermore, it has an unusually high intrinsic X-ray luminosity of $\sim1.6\times10^{30}$~ergs~s$^{-1}$ for a single, late-F-type star \citep{1994A&A...285..272T}.

The first photometric monitoring of the star by \citet{1996A&AS..115...41C} did not reveal photometric variability. However, the flat-bottomed shapes of Stokes I line intensity profiles suggested that a dark spot is present on AF~Lep \citep{2002PASA...19..527B}. Later, a photometric follow-up revealed a prominent photometric minimum in a phase diagram \citep{2003IBVS.5451....1B}.

This paper has two parts. In the first part, we explore what we can learn about AF~Lep via photometric and spectroscopic observations. The second part of the paper consists of theoretical calculations aiming to determine whether high-latitude spots can be formed.


\section{ Observations and data reduction }\label{sec:data}

\subsection{Photometric observations}

\begin{figure} 
\resizebox{\hsize}{!}{\includegraphics{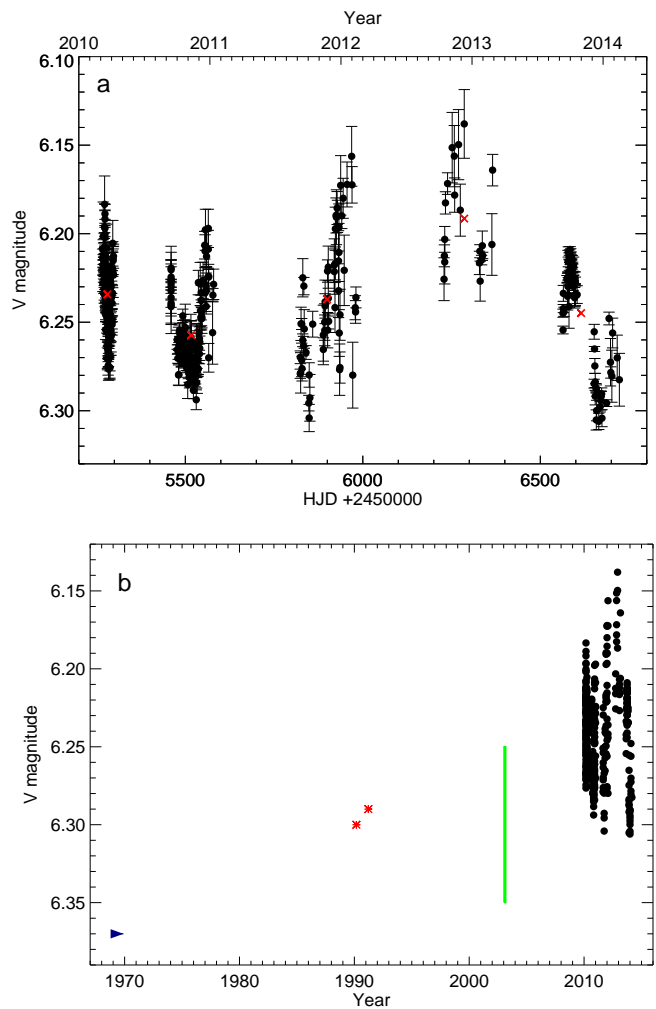}} 
\caption{ {\bf a)} V-band photometric data of AF~Lep with measurement errors used in this paper. The mean value of each observing season is plotted with red x. {\bf b)} The whole photometric record in the V-band: the blue arrowhead represents data from \citet{1972MNRAS.159..165S}, the red asterisks data from \citet{1996A&AS..115...41C}, the green bar illustrates the magnitude range from \citet{2003IBVS.5451....1B}, and black dots represent this work. }
\label{photo}
\end{figure}

We started to observe AF~Lep in March/April 2010 with the Amadeus, a T7 Automatic Photoelectric Telescope (APT) at Fairborn Observatory, jointly operated by the University of Vienna and Leibniz-Institute f\"ur Astrophysik Potsdam (AIP), with Johnson-Cousins V filter (for more details see \citealt{1997PASP..109..697S}). After that, four additional data sets were obtained (see Fig.~\ref{photo}). During the first observing run the star was observed from four to nine times per night for a month. The subsequent runs lasted longer -- from four to five months -- but the star was observed less frequently, and there are only one or two data points per night. Measurements were made differentially between the target, a comparison star (HD~36379), a check star (HD~34538), and the sky position. The data reduction is automated and was described by \citet{1997A&AS..125...11S} and \citet{2001AN....322..325G}. 

\subsection{Spectroscopic and spectropolarimetric observations}

Most of the spectroscopic and spectropolarimetric observations of AF~Lep were carried out during two simultaneous observing runs with the fibre-fed STELLA Echelle Spectrograph (SES) mounted on STELLA-I (Tenerife; \citealt{2004AN....325..527S}; \citealt{2001AN....322..295G}) and the Anglo-Australian Telescope (AAT) with the SEMPOL visiting polarimeter \citep{1993A&A...278..231S, 2011MNRAS.413.1922M} during December 2008 and November/December 2009. The STELLA observations were also continued into January 2010. Furthermore, AF~Lep had been observed earlier in November 2005 with SOFIN at the Nordic Optical Telescope (NOT, La Palma). The STELLA data have a continuous wavelength coverage (3900--9000~\AA) with a resolving power of 55\,000. The spectra from the AAT have a shorter wavelength coverage (4380--6810~\AA) but higher resolution ($\sim$70\,000). The SOFIN \'echelle spectrograph, acquired with the second camera, provides 33 useful orders in a spectral range of 3930--9040~\AA. The slit width of 65~$\mu$m centred at 6427~\AA\ gives a resolution of 76\,000. More information is presented in Tables~\ref{stellaobs}, \ref{aatobs} and \ref{sofinobs}, which are only available online.

\onltab{
\begin{table*}
\caption{The spectroscopic observations of AF~Lep with SES at the STELLA
observatory. The S/N given is an average value for spectral region 4750--6900\AA.}
\label{stellaobs}
\begin{tabular}{cccrr|cccrr}
\hline \hline
HJD  & day & phase & S/N & Exp.\ t & HJD  & day & phase & S/N & Exp.\ t \\
2450000+ & & & & s & 2450000+ & & & & s \\
\hline
\multicolumn{4}{c}{\emph{2008}}        &      & 5168.581883 & 03/12/2009 & 0.85 &  119 & 1200 \\
4806.449041 & 05/12/2008 & 0.97 &   84 & 1200   & 5168.612782 & 03/12/2009 & 0.88 &   92 & 1200 \\
4806.488960 & 05/12/2008 & 0.01 &   69 & 1200   & 5168.651441 & 03/12/2009 & 0.92 &  113 & 1200 \\
4807.442635 & 06/12/2008 & 1.00 &   68 & 1144   & 5168.684299 & 03/12/2009 & 0.95 &   88 & 1200 \\
4807.485140 & 06/12/2008 & 0.04 &   75 & 1200   & 5168.724356 & 03/12/2009 & 1.00 &   71 & 1200 \\
4807.500198 & 06/12/2008 & 0.06 &   71 & 1200   & 5169.677176 & 04/12/2009 & 0.98 &   93 & 1200 \\
4808.455330 & 07/12/2008 & 0.05 &   92 & 1200   & 5169.706822 & 04/12/2009 & 0.01 &   74 & 1200 \\
4808.506302 & 07/12/2008 & 0.10 &   91 & 1200   & \multicolumn{4}{c}{\emph{2010}} \\
4808.686498 & 08/12/2008 & 0.29 &   95 & 1200   & 5210.375680 & 13/01/2010 & 0.11 &  107 & 1200 \\
4809.455377 & 08/12/2008 & 0.08 &   91 & 1200   & 5210.481473 & 13/01/2010 & 0.22 &  113 & 1200 \\
4809.514308 & 09/12/2008 & 0.14 &   79 & 1200   & 5210.510337 & 13/01/2010 & 0.25 &  113 & 1200 \\
\multicolumn{4}{c}{\emph{2009}}        &      & 5210.615744 & 14/01/2010 & 0.36 &   70 & 1200 \\
5163.435205 & 27/11/2009 & 0.52 &   85 & 1200   & 5211.432112 & 14/01/2010 & 0.21 &  123 & 1200 \\
5163.452347 & 27/11/2009 & 0.54 &   81 & 1200   & 5211.468481 & 14/01/2010 & 0.24 &  126 & 1200 \\
5163.485338 & 27/11/2009 & 0.57 &  100 & 1200   & 5211.503232 & 14/01/2010 & 0.28 &  121 & 1200 \\
5163.513160 & 28/11/2009 & 0.60 &   93 & 1200   & 5211.542212 & 15/01/2010 & 0.32 &  136 & 1200 \\
5163.552318 & 28/11/2009 & 0.64 &   92 & 1200   & 5211.570609 & 15/01/2010 & 0.35 &  125 & 1200 \\
5163.580713 & 28/11/2009 & 0.67 &  101 & 1200   & 5211.600474 & 15/01/2010 & 0.38 &   89 & 1200 \\
5163.612288 & 28/11/2009 & 0.70 &  106 & 1200   & 5212.416273 & 15/01/2010 & 0.22 &  121 & 1200 \\
5163.640174 & 28/11/2009 & 0.73 &   85 & 1200   & 5212.456635 & 15/01/2010 & 0.27 &  125 & 1200 \\
5163.672249 & 28/11/2009 & 0.77 &  102 & 1200   & 5212.489709 & 15/01/2010 & 0.30 &  128 & 807 \\
5163.700904 & 28/11/2009 & 0.79 &   90 & 1200   & 5212.566831 & 16/01/2010 & 0.38 &  133 & 1200 \\
5163.736466 & 28/11/2009 & 0.83 &   63 & 1200   & 5212.606046 & 16/01/2010 & 0.42 &   97 & 1200 \\
5164.503297 & 28/11/2009 & 0.63 &   94 & 1200   & 5213.412320 & 16/01/2010 & 0.26 &   99 & 1200 \\
5164.640697 & 29/11/2009 & 0.77 &   86 & 1200   & 5213.455694 & 16/01/2010 & 0.30 &  116 & 1200 \\
5164.674109 & 29/11/2009 & 0.80 &  106 & 1200   & 5213.499864 & 16/01/2010 & 0.35 &  103 & 1200 \\
5165.457987 & 29/11/2009 & 0.61 &   90 & 1200   & 5213.541172 & 17/01/2010 & 0.39 &  109 & 1200 \\
5165.499258 & 29/11/2009 & 0.66 &   97 & 1200   & 5213.569589 & 17/01/2010 & 0.42 &  119 & 1200 \\
5165.529145 & 30/11/2009 & 0.69 &   99 & 1200   & 5214.413196 & 17/01/2010 & 0.29 &   91 & 1200 \\
5165.568430 & 30/11/2009 & 0.73 &   93 & 1200   & 5214.450284 & 17/01/2010 & 0.33 &   92 & 1200 \\
5165.596842 & 30/11/2009 & 0.76 &   76 & 1200   & 5214.484901 & 17/01/2010 & 0.37 &  100 & 1200 \\
5165.625225 & 30/11/2009 & 0.79 &   99 & 1200   & 5214.513877 & 18/01/2010 & 0.40 &   97 & 1200 \\
5165.652971 & 30/11/2009 & 0.82 &   92 & 1200   & 5214.561562 & 18/01/2010 & 0.45 &   80 & 1200 \\
5165.680734 & 30/11/2009 & 0.84 &   97 & 1200   & 5214.594582 & 18/01/2010 & 0.48 &   55 & 1200 \\
5165.718373 & 30/11/2009 & 0.88 &   65 & 1200   & 5215.332572 & 18/01/2010 & 0.24 &   79 &  555 \\
5166.658329 & 01/12/2009 & 0.86 &   67 & 828   & 5215.370475 & 18/01/2010 & 0.28 &   36 &  300 \\
5167.459250 & 01/12/2009 & 0.69 &   88 & 1200   & 5215.482607 & 18/01/2010 & 0.40 &  122 & 1200 \\
5167.487761 & 01/12/2009 & 0.72 &   88 & 1200   & 5215.524929 & 19/01/2010 & 0.44 &  132 & 1200 \\
5167.521360 & 02/12/2009 & 0.75 &  104 & 1200   & 5215.576123 & 19/01/2010 & 0.50 &  126 & 1200 \\
5167.644374 & 02/12/2009 & 0.88 &  104 & 1200   & 5216.414315 & 19/01/2010 & 0.36 &  132 & 1200 \\
5167.686452 & 02/12/2009 & 0.92 &   81 & 1200   & 5216.443441 & 19/01/2010 & 0.39 &  127 & 1200 \\
5167.716547 & 02/12/2009 & 0.95 &   89 & 1200   & 5216.479376 & 19/01/2010 & 0.43 &  109 & 1200 \\
5168.430527 & 02/12/2009 & 0.69 &   78 & 1200   & 5216.520433 & 20/01/2010 & 0.47 &   74 & 1200 \\
5168.463362 & 02/12/2009 & 0.73 &   84 & 1200   & 5216.569482 & 20/01/2010 & 0.52 &  125 & 1200 \\
5168.496477 & 02/12/2009 & 0.76 &   93 & 1200   & 5216.597955 & 20/01/2010 & 0.55 &   96 & 1200 \\
5168.550220 & 03/12/2009 & 0.81 &  117 & 1200   &             &            &      &      &    \\
\hline
\end{tabular}
\end{table*}
}

\onllongtab{
\begin{longtable}{cccrr|cccrr}
\caption{\label{aatobs} The spectroscopic observations of AF~Lep with SemelPol at the AAT. The S/N given is an average value for spectral region 4750--6900\AA.}\\
\hline 
\hline
HJD  & day & phase & S/N & Exp.\ t & HJD  & day & phase & S/N & Exp.\ t \\
2450000+ & & & & s & 2450000+ & & & & s \\
\hline
\endfirsthead
\caption{Continued.} \\
\hline
HJD  & day & phase & S/N & Exp.\ t & HJD  & day & phase & S/N & Exp.\ t \\
2450000+ & & & & s & 2450000+ & & & & s \\
\hline
\endhead
\hline
\endfoot
\hline
\endlastfoot
\multicolumn{4}{c}{\emph{2008}}     &     & 4814.968 & 14/12/2008 & 0.79 &  100 & 400 \\
4806.107 & 05/12/2008 & 0.61 &   25 & 400 & 4814.974 & 14/12/2008 & 0.79 &  100 & 400 \\
4806.113 & 05/12/2008 & 0.62 &   26 & 400 & 4814.979 & 14/12/2008 & 0.80 &  100 & 400 \\
4806.118 & 05/12/2008 & 0.63 &   61 & 400 & 4814.985 & 14/12/2008 & 0.81 &  100 & 400 \\
4806.124 & 05/12/2008 & 0.63 &   32 & 400 & 4815.053 & 14/12/2008 & 0.88 &   96 & 400 \\
4806.130 & 05/12/2008 & 0.64 &   25 & 400 & 4815.059 & 14/12/2008 & 0.88 &  100 & 400 \\
4806.135 & 05/12/2008 & 0.64 &   70 & 400 & 4815.065 & 14/12/2008 & 0.89 &  100 & 400 \\
4806.141 & 05/12/2008 & 0.65 &   60 & 400 & 4815.070 & 14/12/2008 & 0.89 &  100 & 400 \\
4806.147 & 05/12/2008 & 0.66 &   67 & 400 & 4815.124 & 14/12/2008 & 0.95 &  110 & 400 \\
4806.153 & 05/12/2008 & 0.66 &   54 & 400 & 4815.130 & 14/12/2008 & 0.96 &  120 & 400 \\
4806.158 & 05/12/2008 & 0.67 &   26 & 400 & 4815.141 & 14/12/2008 & 0.97 &  100 & 400 \\
4806.164 & 05/12/2008 & 0.67 &   53 & 400 & 4816.078 & 15/12/2008 & 0.94 &  190 & 400 \\
4806.169 & 05/12/2008 & 0.68 &   40 & 400 & 4816.083 & 15/12/2008 & 0.94 &  190 & 400 \\
4806.175 & 05/12/2008 & 0.69 &   26 & 400 & 4816.089 & 15/12/2008 & 0.95 &  190 & 400 \\
4806.181 & 05/12/2008 & 0.69 &   21 & 400 & 4816.095 & 15/12/2008 & 0.95 &  190 & 400 \\
4806.220 & 05/12/2008 & 0.73 &   77 & 400 & 4816.169 & 15/12/2008 & 0.03 &  160 & 400 \\
4806.226 & 05/12/2008 & 0.74 &   69 & 400 & 4816.175 & 15/12/2008 & 0.04 &  160 & 400 \\
4806.254 & 05/12/2008 & 0.77 &   45 & 400 & 4816.180 & 15/12/2008 & 0.04 &  170 & 400 \\
4806.260 & 05/12/2008 & 0.77 &   67 & 400 & 4816.186 & 15/12/2008 & 0.05 &  160 & 400 \\
4808.033 & 07/12/2008 & 0.61 &   75 & 400 & 4817.110 & 16/12/2008 & 0.01 &  180 & 400 \\
4808.039 & 07/12/2008 & 0.61 &  100 & 400 & 4817.115 & 16/12/2008 & 0.01 &  180 & 400 \\
4808.045 & 07/12/2008 & 0.62 &  100 & 400 & 4817.121 & 16/12/2008 & 0.02 &  180 & 400 \\
4808.050 & 07/12/2008 & 0.63 &   29 & 400 & 4817.127 & 16/12/2008 & 0.02 &  180 & 400 \\
4808.055 & 07/12/2008 & 0.63 &   28 & 400 & 4817.178 & 16/12/2008 & 0.08 &  140 & 400 \\
4809.015 & 08/12/2008 & 0.63 &  160 & 400 & 4817.184 & 16/12/2008 & 0.08 &  130 & 400 \\
4809.020 & 08/12/2008 & 0.63 &  140 & 400 & 4817.190 & 16/12/2008 & 0.09 &  130 & 400 \\
4809.026 & 08/12/2008 & 0.64 &  160 & 400 & 4817.195 & 16/12/2008 & 0.09 &  130 & 400 \\
4809.032 & 08/12/2008 & 0.64 &  150 & 400 & 4817.933 & 17/12/2008 & 0.86 &  110 & 400 \\
4809.128 & 08/12/2008 & 0.74 &  160 & 400 & 4817.938 & 17/12/2008 & 0.86 &  120 & 400 \\
4809.134 & 08/12/2008 & 0.75 &  170 & 400 & 4817.944 & 17/12/2008 & 0.87 &  110 & 400 \\
4809.139 & 08/12/2008 & 0.75 &  160 & 400 & 4817.950 & 17/12/2008 & 0.87 &  100 & 400 \\
4809.145 & 08/12/2008 & 0.76 &  170 & 400 & 4818.024 & 17/12/2008 & 0.95 &  130 & 400 \\
4809.197 & 08/12/2008 & 0.81 &  130 & 400 & 4818.041 & 17/12/2008 & 0.97 &  100 & 400 \\
4809.203 & 08/12/2008 & 0.82 &  150 & 400 & 4818.130 & 17/12/2008 & 0.06 &  130 & 400 \\
4809.208 & 08/12/2008 & 0.82 &  140 & 400 & 4818.136 & 17/12/2008 & 0.07 &   94 & 400 \\
4809.214 & 08/12/2008 & 0.83 &  130 & 400 & 4818.141 & 17/12/2008 & 0.07 &  120 & 400 \\
4809.962 & 09/12/2008 & 0.61 &  110 & 400 & 4818.147 & 17/12/2008 & 0.08 &  130 & 400 \\
4809.968 & 09/12/2008 & 0.61 &  120 & 400 & 4818.153 & 17/12/2008 & 0.08 &  150 & 400 \\
4809.974 & 09/12/2008 & 0.62 &  120 & 400 & 4818.158 & 17/12/2008 & 0.09 &  150 & 400 \\
4809.979 & 09/12/2008 & 0.62 &  130 & 400 & 4818.164 & 17/12/2008 & 0.10 &  140 & 400 \\
4810.966 & 10/12/2008 & 0.64 &   95 & 400 & 4818.170 & 17/12/2008 & 0.10 &  130 & 400 \\
4810.971 & 10/12/2008 & 0.65 &  100 & 400 & 4818.913 & 18/12/2008 & 0.87 &   76 & 400 \\
4810.977 & 10/12/2008 & 0.66 &  140 & 400 & 4818.919 & 18/12/2008 & 0.88 &   99 & 400 \\
4810.983 & 10/12/2008 & 0.66 &  110 & 400 & 4818.925 & 18/12/2008 & 0.88 &   88 & 400 \\
4810.988 & 10/12/2008 & 0.67 &  110 & 400 & 4818.930 & 18/12/2008 & 0.89 &   98 & 400 \\
4810.994 & 10/12/2008 & 0.67 &  140 & 400 & 4818.936 & 18/12/2008 & 0.90 &  100 & 400 \\
4811.000 & 10/12/2008 & 0.68 &  140 & 400 & 4818.942 & 18/12/2008 & 0.90 &  120 & 400 \\
4811.005 & 10/12/2008 & 0.69 &  100 & 400 & 4818.947 & 18/12/2008 & 0.91 &  120 & 400 \\
4811.058 & 10/12/2008 & 0.74 &  120 & 400 & 4818.953 & 18/12/2008 & 0.91 &  110 & 400 \\
4811.064 & 10/12/2008 & 0.75 &  150 & 400 & 4818.959 & 18/12/2008 & 0.92 &   50 & 400 \\
4811.070 & 10/12/2008 & 0.75 &  160 & 400 & 4818.964 & 18/12/2008 & 0.92 &   80 & 400 \\
4811.075 & 10/12/2008 & 0.76 &  160 & 400 & 4818.976 & 18/12/2008 & 0.94 &  110 & 400 \\
4811.175 & 10/12/2008 & 0.86 &  180 & 400 & 4818.982 & 18/12/2008 & 0.94 &  120 & 400 \\
4811.180 & 10/12/2008 & 0.87 &  190 & 400 & 4818.987 & 18/12/2008 & 0.95 &  110 & 400 \\
4811.186 & 10/12/2008 & 0.87 &  180 & 400 & 4818.993 & 18/12/2008 & 0.95 &   41 & 400 \\
4811.192 & 10/12/2008 & 0.88 &  170 & 400 & 4818.999 & 18/12/2008 & 0.96 &   61 & 400 \\
4811.260 & 10/12/2008 & 0.95 &  130 & 400 & 4819.005 & 18/12/2008 & 0.97 &  110 & 400 \\
4811.266 & 10/12/2008 & 0.96 &  130 & 400 & 4819.010 & 18/12/2008 & 0.97 &  120 & 400 \\
4811.272 & 10/12/2008 & 0.96 &  110 & 400 & 4819.016 & 18/12/2008 & 0.98 &  140 & 400 \\
4819.021 & 18/12/2008 & 0.98 &  150 & 400 & 5163.975 & 28/11/2009 & 0.08 &  110 & 400 \\
4819.081 & 18/12/2008 & 0.05 &  130 & 400 & 5163.981 & 28/11/2009 & 0.08 &  130 & 400 \\
4819.087 & 18/12/2008 & 0.05 &  150 & 400 & 5163.987 & 28/11/2009 & 0.09 &  120 & 400 \\
4819.093 & 18/12/2008 & 0.06 &  170 & 400 & 5163.992 & 28/11/2009 & 0.10 &  130 & 400 \\
4819.098 & 18/12/2008 & 0.06 &  160 & 400 & 5164.079 & 28/11/2009 & 0.19 &  150 & 400 \\
4819.104 & 18/12/2008 & 0.07 &  160 & 400 & 5164.085 & 28/11/2009 & 0.19 &  170 & 400 \\
4819.110 & 18/12/2008 & 0.08 &  150 & 400 & 5164.091 & 28/11/2009 & 0.20 &  160 & 400 \\
4819.115 & 18/12/2008 & 0.08 &  170 & 400 & 5164.096 & 28/11/2009 & 0.20 &  150 & 400 \\
4819.121 & 18/12/2008 & 0.09 &  160 & 400 & 5164.102 & 28/11/2009 & 0.21 &  150 & 400 \\
4819.174 & 18/12/2008 & 0.14 &  100 & 400 & 5164.108 & 28/11/2009 & 0.22 &  160 & 400 \\
4819.180 & 18/12/2008 & 0.15 &   99 & 400 & 5164.113 & 28/11/2009 & 0.22 &  140 & 400 \\
4819.186 & 18/12/2008 & 0.15 &  100 & 400 & 5164.119 & 28/11/2009 & 0.23 &  160 & 400 \\
4819.191 & 18/12/2008 & 0.16 &  110 & 400 & 5164.205 & 28/11/2009 & 0.32 &  150 & 400 \\
4819.197 & 18/12/2008 & 0.17 &  110 & 400 & 5164.210 & 28/11/2009 & 0.32 &  160 & 400 \\
4819.203 & 18/12/2008 & 0.17 &   99 & 400 & 5164.216 & 28/11/2009 & 0.33 &  130 & 400 \\
4819.208 & 18/12/2008 & 0.18 &  110 & 400 & 5164.221 & 28/11/2009 & 0.33 &  150 & 400 \\
4819.214 & 18/12/2008 & 0.18 &   99 & 400 & 5164.227 & 28/11/2009 & 0.34 &  140 & 400 \\
\multicolumn{4}{c}{\emph{2009}}     &     & 5164.233 & 28/11/2009 & 0.35 &  150 & 400 \\
5160.988 & 25/11/2009 & 0.99 &   96 & 400 & 5164.238 & 28/11/2009 & 0.35 &  150 & 400 \\
5160.994 & 25/11/2009 & 0.99 &  110 & 400 & 5164.244 & 28/11/2009 & 0.36 &  150 & 400 \\
5160.999 & 25/11/2009 & 1.00 &  130 & 400 & 5164.250 & 28/11/2009 & 0.36 &  150 & 400 \\
5161.005 & 25/11/2009 & 0.00 &   89 & 400 & 5164.256 & 28/11/2009 & 0.37 &  130 & 400 \\
5161.011 & 25/11/2009 & 0.01 &  100 & 400 & 5164.261 & 28/11/2009 & 0.37 &  120 & 400 \\
5161.016 & 25/11/2009 & 0.02 &  130 & 400 & 5164.959 & 29/11/2009 & 0.10 &  120 & 400 \\
5161.028 & 25/11/2009 & 0.03 &   92 & 400 & 5164.965 & 29/11/2009 & 0.10 &  130 & 400 \\
5161.074 & 25/11/2009 & 0.08 &  170 & 400 & 5164.971 & 29/11/2009 & 0.11 &  120 & 400 \\
5161.079 & 25/11/2009 & 0.08 &  160 & 400 & 5164.976 & 29/11/2009 & 0.11 &  130 & 400 \\
5161.085 & 25/11/2009 & 0.09 &  170 & 400 & 5164.982 & 29/11/2009 & 0.12 &  110 & 400 \\
5161.091 & 25/11/2009 & 0.09 &  140 & 400 & 5164.988 & 29/11/2009 & 0.13 &  120 & 400 \\
5161.214 & 25/11/2009 & 0.22 &  140 & 400 & 5164.993 & 29/11/2009 & 0.13 &  120 & 400 \\
5161.220 & 25/11/2009 & 0.23 &  140 & 400 & 5164.999 & 29/11/2009 & 0.14 &  110 & 400 \\
5161.225 & 25/11/2009 & 0.23 &  140 & 400 & 5165.086 & 29/11/2009 & 0.23 &  120 & 400 \\
5161.231 & 25/11/2009 & 0.24 &  150 & 400 & 5165.091 & 29/11/2009 & 0.23 &  120 & 400 \\
5162.963 & 27/11/2009 & 0.03 &  150 & 400 & 5165.097 & 29/11/2009 & 0.24 &  120 & 400 \\
5162.969 & 27/11/2009 & 0.04 &  150 & 400 & 5165.103 & 29/11/2009 & 0.25 &  120 & 400 \\
5162.974 & 27/11/2009 & 0.04 &  160 & 400 & 5165.108 & 29/11/2009 & 0.25 &  100 & 400 \\
5162.980 & 27/11/2009 & 0.05 &  160 & 400 & 5165.120 & 29/11/2009 & 0.26 &  120 & 400 \\
5162.986 & 27/11/2009 & 0.05 &  160 & 400 & 5165.125 & 29/11/2009 & 0.27 &  120 & 400 \\
5162.991 & 27/11/2009 & 0.06 &  160 & 400 & 5165.211 & 29/11/2009 & 0.36 &  110 & 400 \\
5162.997 & 27/11/2009 & 0.07 &  160 & 400 & 5165.216 & 29/11/2009 & 0.36 &  100 & 400 \\
5163.003 & 27/11/2009 & 0.07 &  160 & 400 & 5165.222 & 29/11/2009 & 0.37 &  120 & 400 \\
5163.089 & 27/11/2009 & 0.16 &  180 & 400 & 5165.228 & 29/11/2009 & 0.38 &  120 & 400 \\
5163.094 & 27/11/2009 & 0.17 &  170 & 400 & 5165.233 & 29/11/2009 & 0.38 &  120 & 400 \\
5163.100 & 27/11/2009 & 0.17 &  170 & 400 & 5165.239 & 29/11/2009 & 0.39 &  110 & 400 \\
5163.106 & 27/11/2009 & 0.18 &  170 & 400 & 5165.245 & 29/11/2009 & 0.39 &  100 & 400 \\
5163.112 & 27/11/2009 & 0.19 &  180 & 400 & 5165.250 & 29/11/2009 & 0.40 &  110 & 400 \\
5163.117 & 27/11/2009 & 0.19 &  190 & 400 & 5166.048 & 30/11/2009 & 0.22 &   89 & 400 \\
5163.123 & 27/11/2009 & 0.20 &  200 & 400 & 5166.054 & 30/11/2009 & 0.23 &   92 & 400 \\
5163.128 & 27/11/2009 & 0.20 &  210 & 400 & 5166.059 & 30/11/2009 & 0.24 &   94 & 400 \\
5163.213 & 27/11/2009 & 0.29 &  180 & 400 & 5166.065 & 30/11/2009 & 0.24 &   71 & 400 \\
5163.219 & 27/11/2009 & 0.30 &  180 & 400 & 5166.208 & 30/11/2009 & 0.39 &   98 & 400 \\
5163.225 & 27/11/2009 & 0.31 &  180 & 400 & 5166.214 & 30/11/2009 & 0.40 &  100 & 400 \\
5163.230 & 27/11/2009 & 0.31 &  180 & 400 & 5166.219 & 30/11/2009 & 0.40 &  110 & 400 \\
5163.236 & 27/11/2009 & 0.32 &  170 & 400 & 5166.225 & 30/11/2009 & 0.41 &  110 & 400 \\
5163.242 & 27/11/2009 & 0.32 &  170 & 400 & 5166.231 & 30/11/2009 & 0.41 &  100 & 400 \\
5163.247 & 27/11/2009 & 0.33 &  170 & 400 & 5166.987 & 01/12/2009 & 0.20 &  170 & 400 \\
5163.253 & 27/11/2009 & 0.33 &  170 & 400 & 5166.993 & 01/12/2009 & 0.20 &  170 & 400 \\
5163.958 & 28/11/2009 & 0.06 &  100 & 400 & 5166.999 & 01/12/2009 & 0.21 &  170 & 400 \\
5163.964 & 28/11/2009 & 0.07 &  110 & 400 & 5167.004 & 01/12/2009 & 0.21 &  170 & 400 \\
5163.970 & 28/11/2009 & 0.07 &  110 & 400 & 5167.010 & 01/12/2009 & 0.22 &  160 & 400 \\
5167.016 & 01/12/2009 & 0.23 &  160 & 400 & 5168.247 & 02/12/2009 & 0.50 &  170 & 400 \\
5167.021 & 01/12/2009 & 0.23 &  150 & 400 & 5168.252 & 02/12/2009 & 0.51 &  160 & 400 \\
5167.027 & 01/12/2009 & 0.24 &  150 & 400 & 5168.258 & 02/12/2009 & 0.51 &  160 & 400 \\
5167.120 & 01/12/2009 & 0.33 &  150 & 400 & 5168.966 & 03/12/2009 & 0.25 &  100 & 400 \\
5167.126 & 01/12/2009 & 0.34 &  150 & 400 & 5168.972 & 03/12/2009 & 0.25 &  110 & 400 \\
5167.131 & 01/12/2009 & 0.35 &  150 & 400 & 5168.978 & 03/12/2009 & 0.26 &  110 & 400 \\
5167.137 & 01/12/2009 & 0.35 &  150 & 400 & 5168.983 & 03/12/2009 & 0.26 &  110 & 400 \\
5167.143 & 01/12/2009 & 0.36 &  140 & 400 & 5168.989 & 03/12/2009 & 0.27 &  120 & 400 \\
5167.148 & 01/12/2009 & 0.36 &  120 & 400 & 5168.995 & 03/12/2009 & 0.28 &  130 & 400 \\
5167.154 & 01/12/2009 & 0.37 &  110 & 400 & 5169.000 & 03/12/2009 & 0.28 &  130 & 400 \\
5167.159 & 01/12/2009 & 0.37 &  110 & 400 & 5169.006 & 03/12/2009 & 0.29 &  130 & 400 \\
5167.248 & 01/12/2009 & 0.47 &  110 & 400 & 5169.094 & 03/12/2009 & 0.38 &  160 & 400 \\
5167.254 & 01/12/2009 & 0.47 &  110 & 400 & 5169.099 & 03/12/2009 & 0.38 &  170 & 400 \\
5167.259 & 01/12/2009 & 0.48 &  120 & 400 & 5169.105 & 03/12/2009 & 0.39 &  170 & 400 \\
5167.265 & 01/12/2009 & 0.48 &  120 & 400 & 5169.111 & 03/12/2009 & 0.40 &  170 & 400 \\
5167.970 & 02/12/2009 & 0.21 &  150 & 400 & 5169.116 & 03/12/2009 & 0.40 &  140 & 400 \\
5167.976 & 02/12/2009 & 0.22 &  150 & 400 & 5169.122 & 03/12/2009 & 0.41 &  130 & 400 \\
5167.981 & 02/12/2009 & 0.23 &  160 & 400 & 5169.128 & 03/12/2009 & 0.41 &  130 & 400 \\
5167.987 & 02/12/2009 & 0.23 &  160 & 400 & 5169.133 & 03/12/2009 & 0.42 &  140 & 400 \\
5168.072 & 02/12/2009 & 0.32 &  150 & 400 & 5169.226 & 03/12/2009 & 0.51 &  110 & 400 \\
5168.078 & 02/12/2009 & 0.33 &  150 & 400 & 5169.231 & 03/12/2009 & 0.52 &  110 & 400 \\
5168.083 & 02/12/2009 & 0.33 &  150 & 400 & 5169.237 & 03/12/2009 & 0.53 &  150 & 400 \\
5168.089 & 02/12/2009 & 0.34 &  160 & 400 & 5169.243 & 03/12/2009 & 0.53 &  140 & 400 \\
5168.095 & 02/12/2009 & 0.34 &  160 & 400 & 5169.249 & 03/12/2009 & 0.54 &  130 & 400 \\
5168.100 & 02/12/2009 & 0.35 &  150 & 400 & 5169.254 & 03/12/2009 & 0.54 &  140 & 400 \\
5168.106 & 02/12/2009 & 0.36 &  150 & 400 & 5169.260 & 03/12/2009 & 0.55 &  140 & 400 \\
5168.112 & 02/12/2009 & 0.36 &  150 & 400 & 5169.265 & 03/12/2009 & 0.56 &  110 & 400 \\
5168.241 & 02/12/2009 & 0.49 &  170 & 400 &          &            &      &      &     \\
\end{longtable}
}

\onltab{
\begin{table*}
\caption{The spectroscopic observations of AF~Lep with SOFIN at the Nordic Optical Telescope. The S/N given is for spectral region around 6430~\AA.}
\label{sofinobs}
\begin{tabular}{cccrr|cccrr}
\hline \hline
HJD  & day & phase & S/N & Exp.\ t & HJD  & day & phase & S/N & Exp.\ t \\
2450000+ & & & & s & 2450000+ & & & & s \\
\hline
\multicolumn{4}{c}{\emph{2005}}       &    & 3692.770714 &  18/11/2005 & 0.09 & 202 &  331 \\
3684.618998 & 10/11/2005 & 0.65 & 277 & 907 & 3693.576113 &  19/11/2005 & 0.93 & 296 & 600 \\
3685.665527 & 11/11/2005 & 0.74 & 190 & 900 & 3693.582053 &  19/11/2005 & 0.93 & 132 & 167 \\
3685.673962 & 11/11/2005 & 0.75 & 127 & 300 & 3695.574304 &  21/11/2005 & 1.00 & 219 & 600 \\
3687.697405 & 13/11/2005 & 0.84 & 179 & 900 & 3695.582746 &  21/11/2005 & 0.00 & 257 & 600 \\
3687.709327 & 13/11/2005 & 0.85 & 207 & 900 & 3695.773502 &  21/11/2005 & 0.20 & 178 & 420 \\
3691.572759 & 17/11/2005 & 0.85 & 267 & 676 & 3695.779279 &  21/11/2005 & 0.20 & 155 & 317 \\
3692.584815 & 18/11/2005 & 0.90 & 254 & 651 & 3696.641835 &  22/11/2005 & 0.10 & 229 & 780 \\
3692.764852 & 18/11/2005 & 0.09 & 259 & 423 & 3696.652217 &  22/11/2005 & 0.11 & 231 & 754 \\
\hline
\end{tabular}
\end{table*}
}


\section{ Determining the rotation period }\label{sec:peri}

As summarised by \citet{2008ApJ...683L.179S}, for example, there are two methods that are used to determine a rotation period of a star. One uses rotational broadening of spectral lines. This method, however, is usable only for stars that rotate faster than a threshold velocity set by the spectral resolution and has always $\sin i$ uncertainty, $i$ being the inclination of a star. Furthermore, it gives an average of the periods of the whole stellar disk. The other method uses the periodic modulation of the stellar flux due to the co-rotating dark and bright features on the stellar surface. The advantages of this method are that it can also be used for stars with long periods, it is only mildly affected by latitudinal differential rotation, and the obtained period corresponds to the period of the latitudes where these features are. However, normally, the information on the latitudes of stellar active regions cannot be extracted from a one-dimensional data set, that is, a light-curve. Lately, starspot detections during planetary transits are also used for estimating rotation periods that represent periods at specific latitudes (see \citealt{2008ApJ...683L.179S}, a.o.).

\citet{2011IAUS..273...89L} discussed differences in the spot distribution of the Sun and active stars. On the Sun, the active regions are a mixture of cool spots and bright faculae that evolve on different timescales. However, in very active stars, the cool spots probably dominate the variation. And not only that, but the optical variability of the Sun is dominated by several active regions at the same time. One has to keep in mind that sunspots and solar active regions have typical lifetimes of days to weeks \citep{2003A&ARv..11..153S}, whereas starspots have been observed to persist over a long time (e.g., \citealt{2009A&ARv..17..251S}).

Co-rotating spots on the surface, pulsations, spot evolution, instrumental effects, and a combination of all of them can introduce periodic variations in the light curves. Despite these effects, light curves often show notable stability in the phase, shape, and amplitude of photometric variability over long times. \citet{2011ApJ...733..115M} discussed the two possible explanations for this: either the spots are longer-lived than the sunspots, or they tend to emerge non-uniformly at preferred long-lived active longitudes. The outcome from the first scenario was that although on the Sun the lifetime of the spots increases linearly with area, and the solar log-normal spot distribution is strongly dominated by small spots, there is no evidence that the log-normal distribution holds for other stars. However, on other stars the small spots dominate as well, although they are very often unresolved (see, e.g., \citealt{2009A&A...493..193L}; \citealt{2010A&A...510A..25S}). Furthermore, lifetimes of individual starspots can be short \citep{2009A&A...506..245M}, but the large-scale spot groups can live long \citep{2002AN....323..349H}. The second possibility was that starspots emerge at preferred longitudes. Here it is assumed that the spots seen in photometry and in Doppler images are not huge individual spots, but rather spot groups consisting of a range of smaller spots emerging at preferred longitudes. This could lead to obtaining observed spottedness over longer periods of time although individual spots come and go. As \citet{2011ApJ...733..115M} stated, the only requirement is that the emergence rate is high enough to maintain a rather stable spot coverage, but this is expected for young and active stars.

The rotation period of AF~Lep has commonly been reported to be around one day: \citet{2003IBVS.5451....1B} used P=1\fd0 for phasing photometry and \citet{2006ASPC..358..401M} reported a period of $\sim$1\fd0 from spectroscopy). This is notoriously near the typical window of ground observations ($P_0=0\fd9972$), which creates challenges for determining the period from ground-based photometry. To achieve a good phase coverage using a single telescope, observations covering a long time-base are needed, which in turn may mean that the surface features (and thus the light curve) have undergone significant changes during this time. We can also expect significant problems with spurious periods very near the real one. 

\begin{figure}
\resizebox{\hsize}{!}{\includegraphics{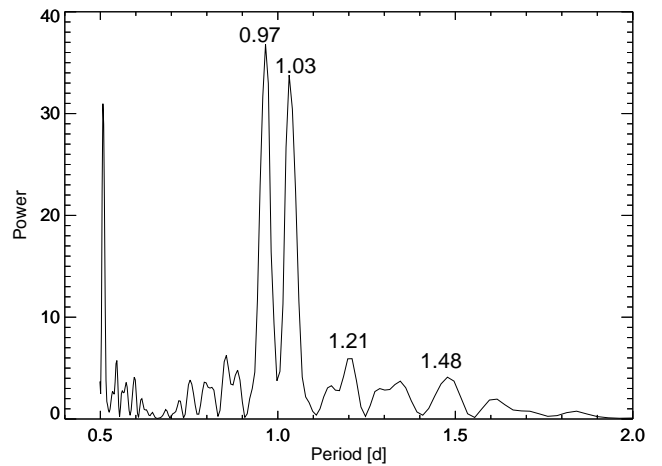}} 
\caption{ Lomb-periodogram showing possible rotation periods. }
\label{pow}
\end{figure}

\begin{figure} 
\resizebox{\hsize}{!}{\includegraphics{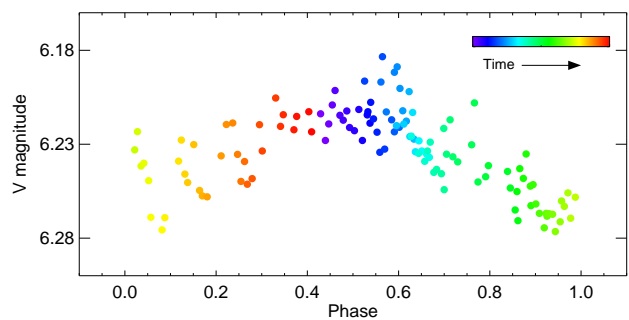}} 
\caption{ Phase diagram for the period $P_{\rm{phot}}=0.9660$ using  the March/April 2010 data set. The observations cover 30 days and the data points have been colour-coded with increasing time of observations.}
\label{afpha}
\end{figure}

We applied time-series analysis methods to the March/April 2010 photometry set (Fig.~\ref{photo}a). Using the Lomb method for unevenly sampled data \citep{1992nrfa.book.....P}, we detected a prominent double peak corresponding to periods of $0\fd97$ and $1\fd03$ (Fig.~\ref{pow}). Using the method described by \citet{1999A&AS..139..629J}, we estimated that the probability of $1\fd03$ being a real period was $ 4 \times 10^{-6}$, while $0\fd97$ passed the test of being non-spurious. This method uses bootstrapping for the model parameter error estimates, evaluates the modelling statistics with the Kolmogorov–Smirnov test, and identifies spurious periodicities from real ones with the phase residual regression. We then applied the continuous-period search method \citep{2011A&A...527A.136L}, which accounts for changing spot patterns, using a window length ($\Delta T_{\max}$) covering the whole data set. In practice, this meant fitting the model
\begin{equation}
y_\mathrm{cps}(t_i,\bar{\beta}) = a_0 + 
\sum_{k=1}^K{[a_k\cos{(2\pi k ft_i)} + b_k\sin{(2\pi k ft_i)}]}
\label{cpsmodel}
\end{equation}
to the observations ($y$). Here $a_0, a_k, b_k$ and $f= 1/P_{\mathrm {phot}}$ are free parameters and $t_i$ the times of the observations. With the optimal order $K=1$, we obtained the period $P_{\mathrm {phot}} \approx 0\fd9660 \pm 0\fd0023$ and the time of the first photometric minimum $\mathrm{HJD}_0=2455268.227 \pm 0.039$. Thus we used the ephemeris
\begin{equation}
\mathrm{HJD}_{\min} = 2455268.227  +0.9660 E
\label{ephem},
\end{equation}
to derive the rotation phases for our photometric and spectroscopic data. The phase diagram using this ephemeris for the March/April 2010 photometry is displayed in Fig.~\ref{afpha}.

\section{ Photometric variability }\label{sec:phovar}

The collected photometry shows both long- and short-term variability. Moreover, if we compare the measured values presented in Fig.~\ref{photo}a with the first reported $V$ magnitude of 6.37 \citep{1972MNRAS.159..165S}, one can conclude that the spot coverage has been larger in the past. The observed magnitudes can also be compared with magnitudes from the early 90s ($V=$6.30 and 6.29; \citealt{1996A&AS..115...41C}), which were on the level of our faintest data points. The photometry from 2003 ($V=$6.35--6.25; \citealt{2003IBVS.5451....1B}) shows that the mean brightness has slowly increased over the past 40 years to the values presented here. The whole photometric record (Fig.~\ref{photo}b) implies that AF~Lep might have a long-term trend, and our four years of observations indicate that there is also variability on shorter timescales, although the data set is not long enough to obtain a reliable cycle length estimate. However, a local minimum occurred around mid-2011, a local maximum was reached around 2012/2013, and in early 2014 the star was dimmer again. Furthermore, the amplitude of the brightness variations implies rotational modulation due to changing spot coverage.

\begin{figure*} 
\resizebox{\hsize}{!}{\includegraphics{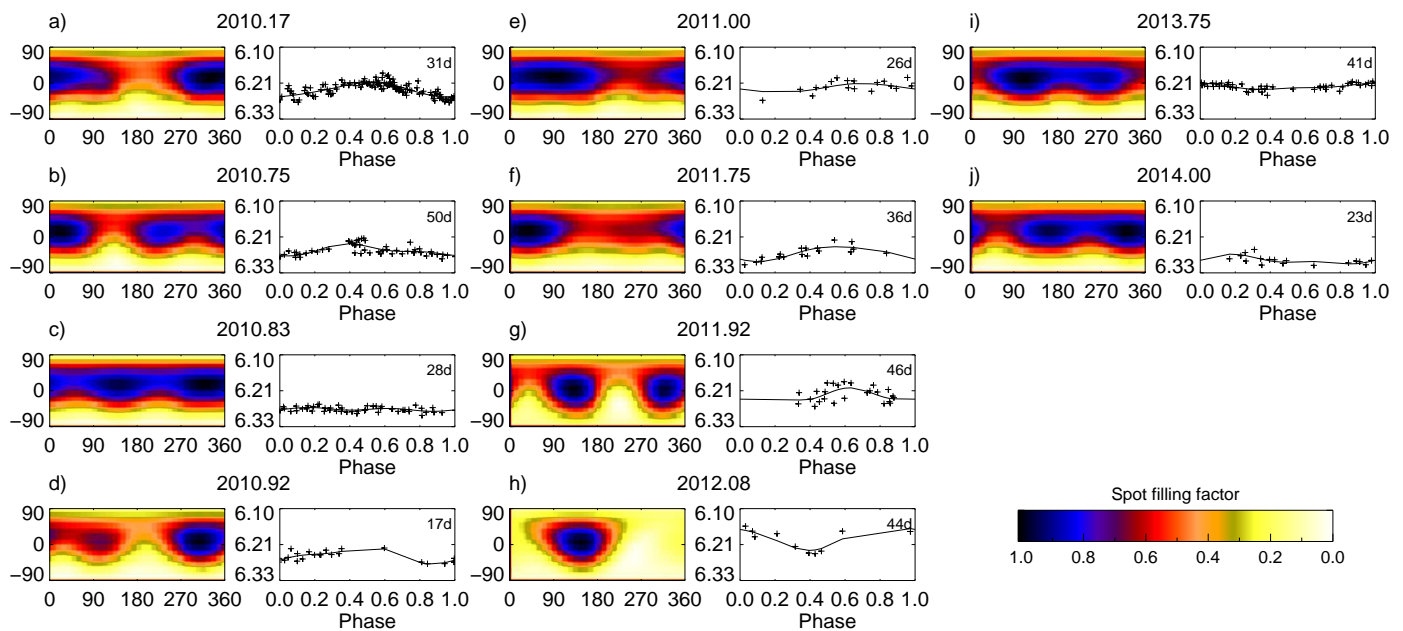}} 
\caption{ Each data subset (a-j) is visualised with two images. The given times represent the middle point of each set. \emph{On the left} we show the light-curve inversion result. The spot-filling factor is larger in the darker regions. A light curve represents a one-dimensional time series and, therefore, the resulting stellar image contains information on the spot distribution only in longitudinal direction. \emph{On the right}, the observed and calculated $V$-band magnitudes are plotted with crosses and lines, respectively. The length of the set (in days) is given in the upper right corner. }
\label{photomaps}
\end{figure*}

To analyse the short-term variability in detail, we inverted the light curves into stellar images (for details about the method, see \citealt{2002A&A...394..505B}). As seen from Figs.~\ref{afpha} and \ref{photomaps}a, from March/April 2010 it is possible to obtain a smooth light curve, where the data points from the beginning of the observing season match those obtained at the end of the season with the obtained period. However, the amplitude of the night-to-night variations is relatively large. In this subset, the spots are concentrated around phase $\varphi=0.0$. However, the second data set (August 2010 -- January 2011), shows significant short-term variations and can be divided into four subsets (Fig~\ref{photomaps}b-e). These sets show moderate phase migration, and a secondary minimum appears in the light curves (Figs.~\ref{photomaps}b-d). A new subset was started when a clear change in the photometric behaviour was detected, that is, a real break in the observations, or when the subsequent data points started to form a maximum instead of a minimum. Variability is also present in later data sets, although observations are rather infrequent and, therefore, it is not always possible to create a continuous light curve over all phases (see Fig.~\ref{photomaps}g), where observations covering 46 days have large scatter and a gap of 0.4 in phase. The photometric minimum is often rather broad, which agrees with high-latitude spots (see Sect.~\ref{sec:di}) and the inclination of the star. 

The drifting of the spot longitudes around $\varphi=0.0$ may be due to an error in the period, which is not easy to determine in this case, as discussed in Sect.~\ref{sec:peri}. An alternative explanation might be a significant surface differential rotation. Using Zeeman-Doppler imaging with Stokes $V$-profiles, \citet{2006ASPC..358..401M} derived values of $\Omega_{\mathrm eq}=6.495\pm 0.011$~rad~d$^{\mathrm -1}$ and $\delta\Omega=0.259\pm 0.019$~rad~d$^{\mathrm -1}$ for AF~Lep. The third option might be drifting due to an azimuthal dynamo wave \citep{2014ApJ...780L..22C}, although we recall that it is a competitive mechanism to the differential rotation, and the drift could be due to a combination of them.

\section{ Surface spot configuration }\label{sec:di}

The spectroscopic observations were phased with the derived new ephemeris (Eq. \ref{ephem}). Local line profiles were calculated with the code of \citet{1991BCrAO..83...89B}. This includes calculating the opacities in the continuum and in atomic and molecular lines, although in this paper molecular lines were omitted. Atomic line parameters were obtained from the Vienna Atomic Line Database (VALD; \citealt{1995A&AS..112..525P,1999A&AS..138..119K}). The stellar model atmospheres used here are from \citet{1993KurCD..13.....K}. The local line profiles were calculated for 20 values of $\mu=\cos \theta$ from the disk centre to the limb. The spectra were calculated for temperatures ranging from 4000~K to 6500~K in steps of 250~K. The Occamian approach was used to invert of the observed line profiles into stellar images \citep{1998A&A...338...97B}. A $6\degr\times6\degr$ grid on the stellar surface was used to integrate local line profiles into normalised flux profiles. With a set of stellar atmosphere models, the stellar image is considered as the distribution of the effective temperature across the stellar surface, as is commonly done in Doppler imaging. This code does not take differential rotation into account. The stellar parameters used are presented in Table~\ref{paramtable}.

\begin{table}
\caption{Stellar parameters of AF~Lep with references.}
\label{paramtable}
\centering
\begin{tabular}{lcc}
\hline \hline
Parameter    & Value          & Ref. \\
\hline
Sp.\ type    & F8~V           & 1 \\
T$_{eff}$     & 6100~K         & 2 \\
$\log$ g     & 4.4            & 2 \\
$\log$ n(Li) & 3.2            & 2 \\
B-V          & 0.55           & 2,3 \\
$v \sin i$   & 50~km~s$^{-1}$  & 2 \\
inclination  & $50\degr$      & 4 \\
Mass         & 1.15~$M_{\sun}$ & 5 \\
Radius       & 1.18~$R_{\sun}$ & 6 \\
Period       & $0\fd9660$~d   & 7 \\
\hline
\end{tabular}
\tablebib{
(1)~\citet{1986AJ.....92..910E}; (2)~\citet{1994A&A...285..272T}; (3)~\citet{1996A&AS..115...41C}; (4)~\citet{2006ASPC..358..401M}; (5)~\citet{1996ApJ...457..340K}; (6)~\citet{1994A&A...282..899B}; (7)~this work.
}
\end{table}

The noise in both STELLA and AAT data was reduced using the least-squares deconvolution (LSD) technique (see \citealt{1989A&A...225..456S} and \citealt{1997MNRAS.291..658D} for the algorithm, and \citealt{2010A&A...521A..86J} for the application). For this purpose, a line mask containing 93 spectral lines was built using lines within a wavelength range of 4770--6900\,\AA. They are mainly \ion{Fe}{i}, \ion{Ni}{i}, and \ion{Ca}{i} lines. This gives a boost from S/N=100 to S/N=230 and from S/N=150 to S/N=320. From the SOFIN data, the surface temperature map was inverted using three spectral lines: \ion{Fe}{i} 6411.64~\AA, \ion{Fe}{i} 6430.80~\AA, and \ion{Ca}{i} 6439.08~\AA.

\begin{figure*} 
\resizebox{\hsize}{!}{\includegraphics{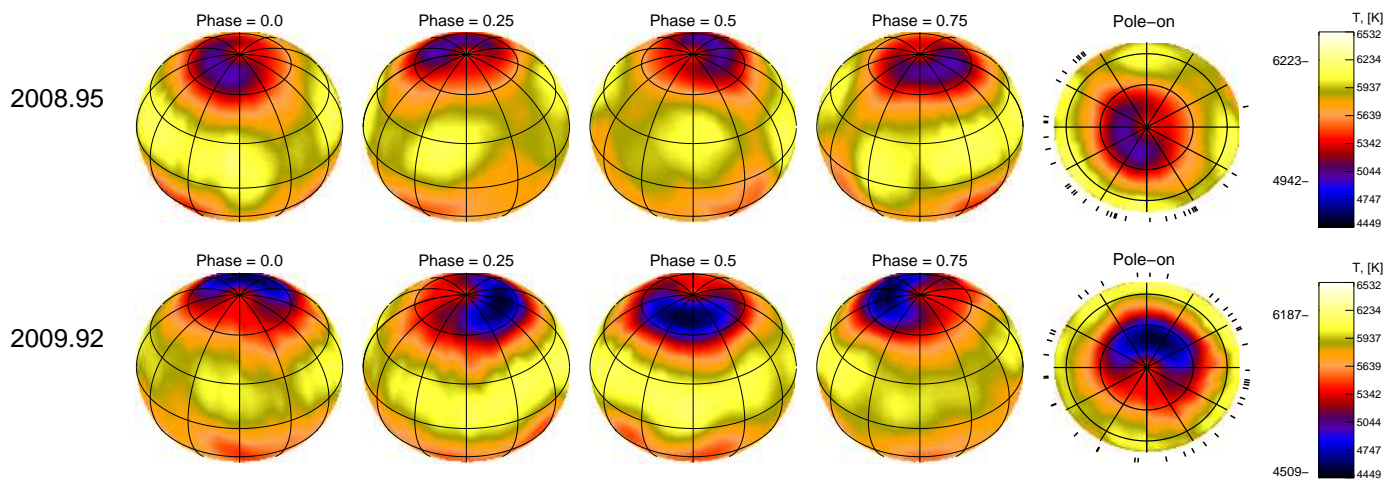}} 
\caption{ Temperature maps of AF~Lep from the data sets with a good phase coverage. They were obtained by combining simultaneous AAT and STELLA observations. For both seasons the map is shown from four angles and pole-on. The used phases have been marked with ticks around the pole-on projection. All maps have a common temperature scale, and the highest and lowest temperature of each map is shown at the left side of the temperature scale.}
\label{dimaps}
\end{figure*}

\begin{figure*} 
\resizebox{\hsize}{!}{\includegraphics{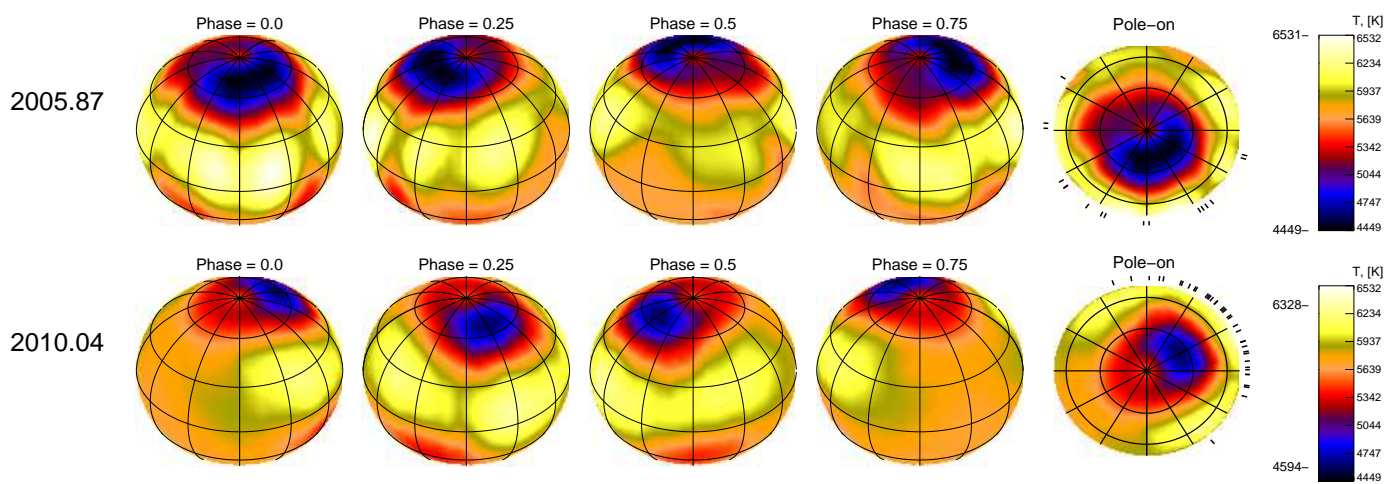}} 
\caption{ Temperature maps of AF~Lep from the data sets with an incomplete phase coverage. The 2005 data set was obtained at the NOT, the 2010 data set with STELLA. Otherwise maps as in Fig.~\ref{dimaps}. }
\label{dimaps2}
\end{figure*}

The resulting temperature maps are shown in Figs.~\ref{dimaps} and \ref{dimaps2}. The fits to the observed spectra and LSD profiles are shown in Figs.~\ref{l2005}--\ref{llsd2010} and the differences between the mean profile and the observed ones are illustrated in Figs.~\ref{tl2005}--\ref{tl2010} to visualise spots moving through the line profiles (both sets of figures are available in electronic form only). To create the temperature map from 2005 observations, all available individual SOFIN observations were used. For the other temperature maps, mainly the SEMPOL spectra from the AAT were used because it gives the best phase coverage, and the long phase-gaps were filled using the STELLA spectra. For the spectropolarimetric SEMPOL observations there are four consecutive Stokes~I exposures for each Stokes~V exposure that were added together. The photospheric temperature of 6200~K agrees with earlier results and the spectral type of the star. A cool spot is located near the polar region of the star. The time separation of the maps in Fig.~\ref{dimaps} is one year.  Although the total spot coverage has seemingly changed very little, clear evolution is detected. In the first map (2008.95) the spot is formed from two components at $\varphi=0.77$ (secondary) and at $\varphi=0.96$ (primary). In the latter map (2009.92), the spot has grown larger and the two components now have equal sizes and form a larger spot that covers half of the polar longitudes. The middle point is at $\varphi=0.47$ and the spot is also cooler. The mean latitude of the spot is around +65\degr. 

The maps presented in Fig.~\ref{dimaps2} have to be interpreted with care because of the large phase gaps in both data sets -- in the first map phases $\varphi=$0.65--0.21 are covered, and the later map covers only phases $\varphi=$0.11--0.55. In the 2005.87 map the main feature is at $\varphi=0.00$ with a tail, which has a mid-point at $\varphi=0.18$. In the 2010.04 map the spot is again smaller with the mean phase of $\varphi=0.35$. This map also has the lowest spot latitude (+61\degr). However, when taking into account the spatial resolution of the maps, the mean latitude of the spots remains relatively constant. The maps inverted from the data with an incomplete phase coverage have a broader temperature range than the maps inverted from the data with a good phase coverage.

The high-latitude spot is anti-symmetric around the polar region in all the maps, and there are no low-latitude features. In this sense the maps presented here differ from the ($\sim$2002.00) map by \citet{2006ASPC..358..401M}, which shows a very compact polar (completely above latitude +75\degr) spot and some very weak low-latitude features.

\onlfig{
\begin{figure}
\resizebox{.5\hsize}{!}{\includegraphics{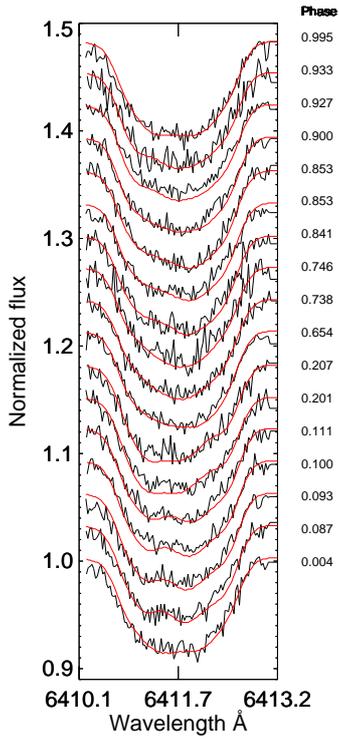}} 
\caption{ Calculated (red line) and observed (black line) spectral lines for 2005.87 data set using \ion{Fe}{i} 6411.64~\AA\ line.}
\label{l2005}
\end{figure}
}

\onlfig{
\begin{figure}
\resizebox{\hsize}{!}{\includegraphics{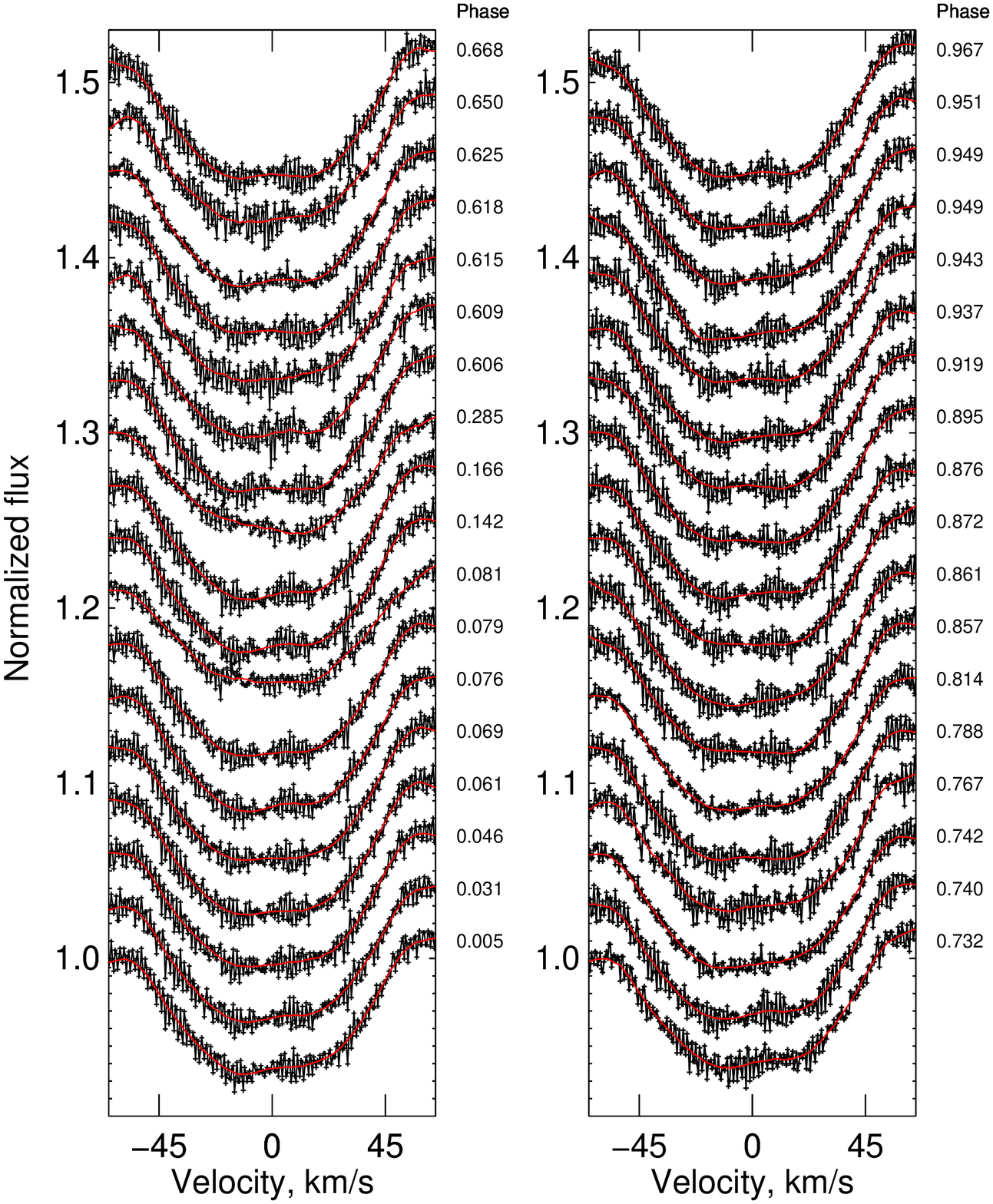}} 
\caption{ Calculated (red line) and observed (black line) LSD profiles for 2008.95 data set.}
\label{llsd2008}
\end{figure}
}

\onlfig{
\begin{figure}
\resizebox{\hsize}{!}{\includegraphics{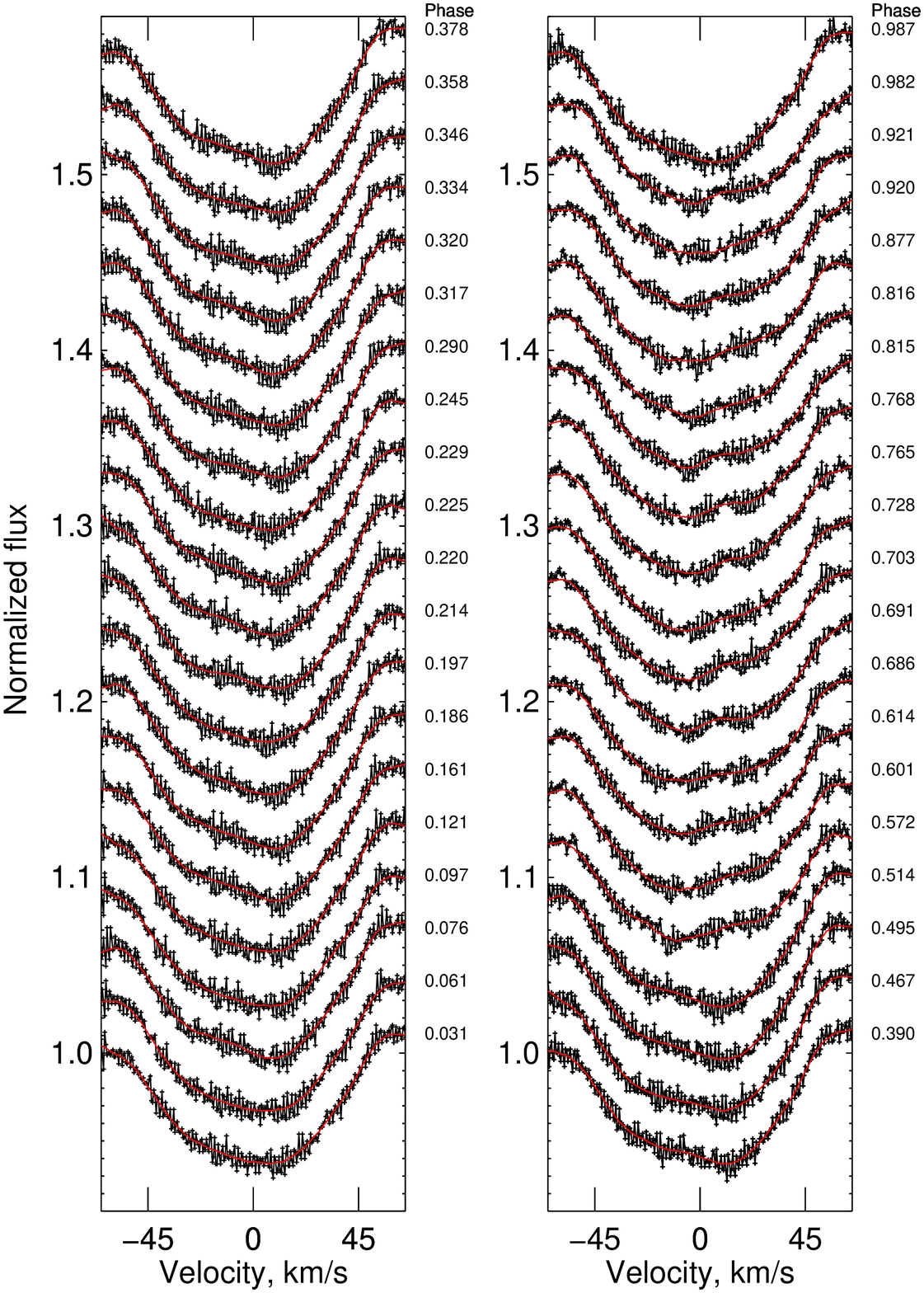}} 
\caption{ Calculated (red line) and observed (black line) LSD profiles for 2009.92 data set.}
\label{llsd2009}
\end{figure}
}

\onlfig{
\begin{figure}
\resizebox{\hsize}{!}{\includegraphics{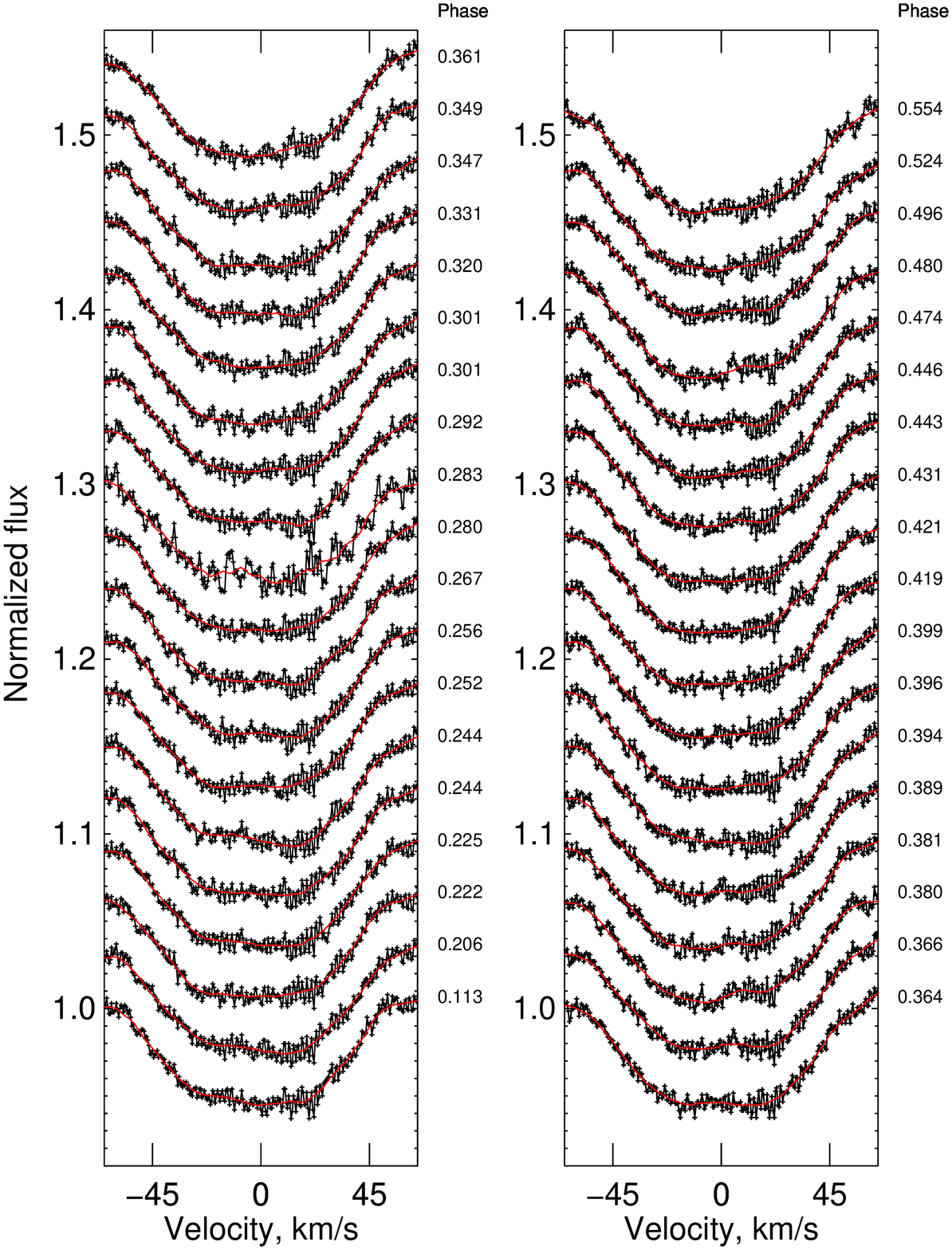}} 
\caption{ Calculated (red line) and observed (black line) LSD profiles for 2010.04 data set.}
\label{llsd2010}
\end{figure}
}

\onlfig{
\begin{figure}
\resizebox{\hsize}{!}{\includegraphics{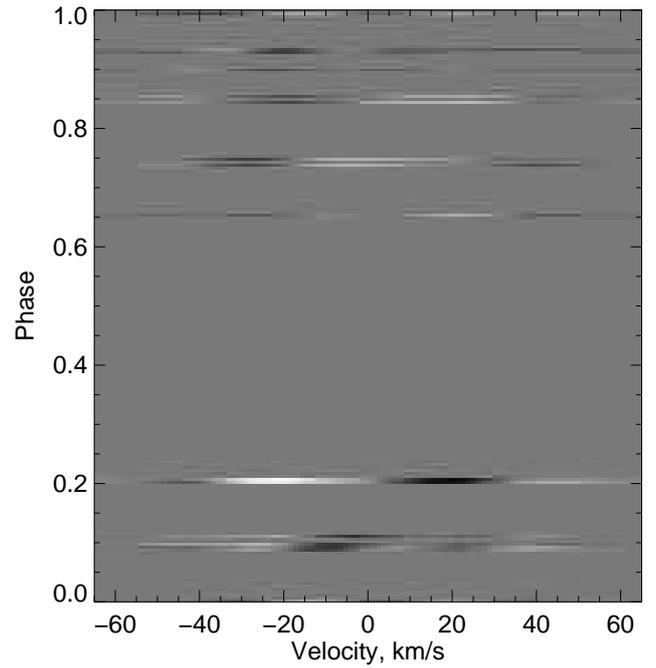}} 
\caption{ The difference between the mean profile and the observed profile for 2005 data to show the movements of the spots through line profiles as a function of phase. The darker colour indicates the presence of a spot.}
\label{tl2005}
\end{figure}
}

\onlfig{
\begin{figure}
\resizebox{\hsize}{!}{\includegraphics{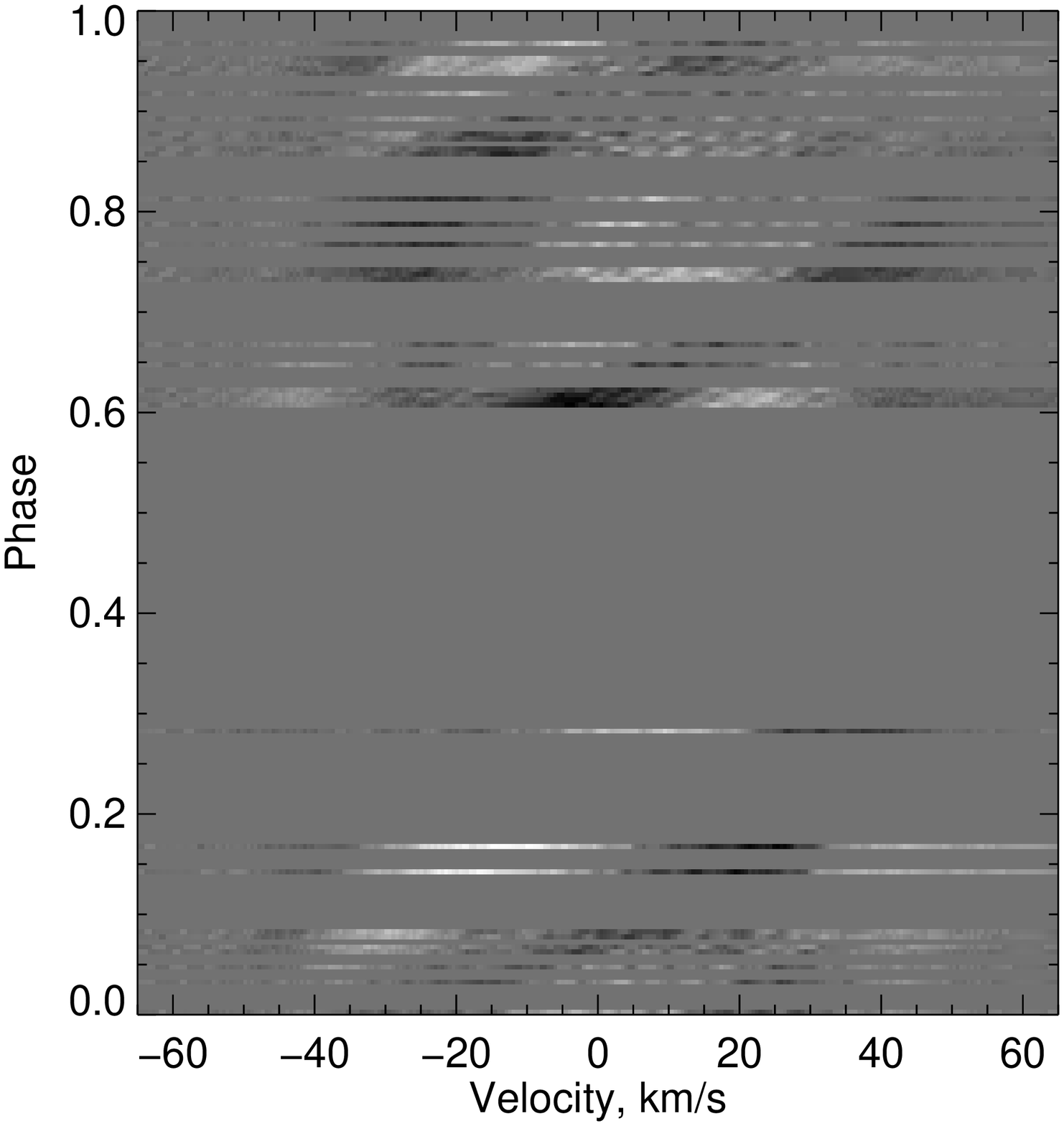}} 
\caption{ As figure \ref{tl2005} but for 2008 data.}
\label{tl2008}
\end{figure}
}

\onlfig{
\begin{figure}
\resizebox{\hsize}{!}{\includegraphics{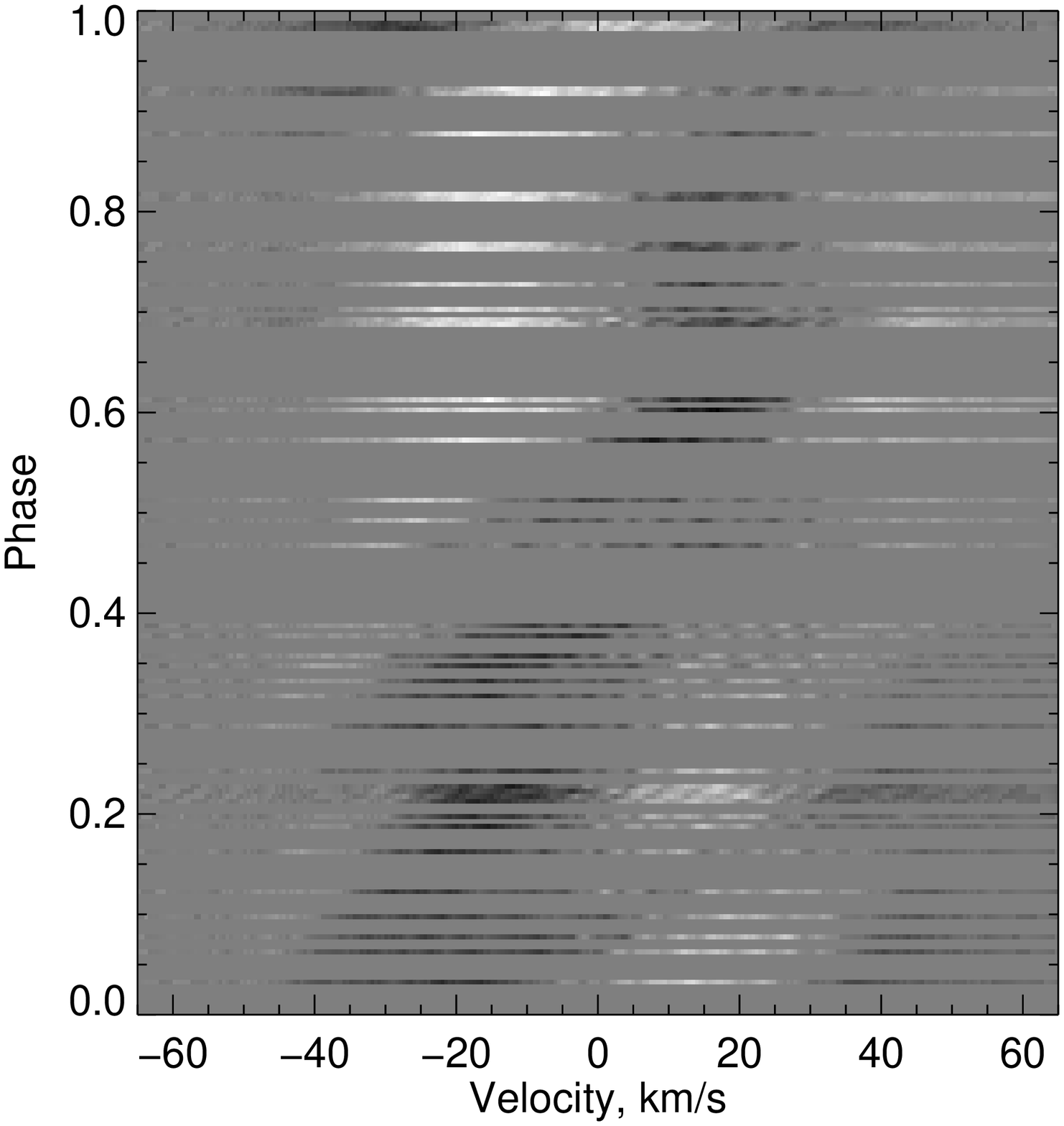}} 
\caption{ As figure \ref{tl2005} but for 2009 data. }
\label{tl2009}
\end{figure}
}

\onlfig{
\begin{figure}
\resizebox{\hsize}{!}{\includegraphics{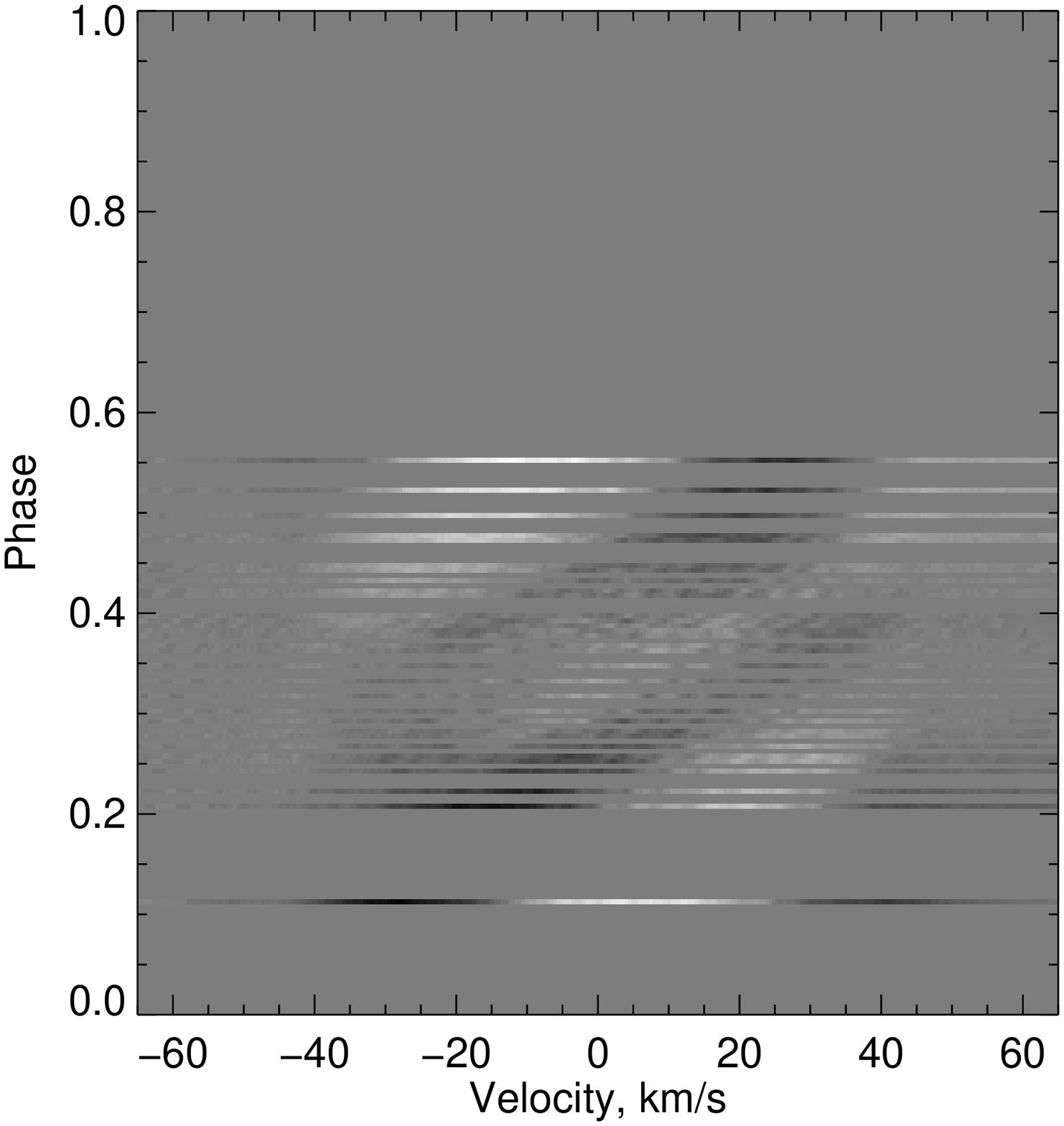}} 
\caption{ As figure \ref{tl2005} but for 2010 data.}
\label{tl2010}
\end{figure}
}

To validate the obtained results, the spot configuration on AF~Lep should be compared with spot configurations on other F dwarfs and with spot configurations on other fast-rotating (P$\le$1~d) dwarfs. Unfortunately, there are not many F dwarfs that have been mapped to date. A short overview of the current knowledge about the spot configuration in comparable stars is presented in Table~\ref{rescomps}. Some binary components are included as well, although they do not make good comparisons because they are very different entities. More comparison targets can be found in an extensive list of \citet{2009A&ARv..17..251S}. 

\begin{table}
\caption{Overview of spot latitudes in comparable targets sorted according to the spectral type of the stars. In the ``Spots'' column P=polar, H=high-latitude, M=mid-latitude, and L=low-latitude. Parenthesis around the letter are used when spots are not always present in that latitude range or are weak. The binarity is stated in the ``Note'' column with ``B''.}
\label{rescomps}
\centering
\begin{tabular}{lllccc}
\hline \hline
Star    & Sp.\ type  & P$_{\rm rot}$  & Spots & Note & Ref. \\
\hline
$\tau$~Boo\tablefootmark{a} & F7~V & $3\fd31$ & L-H &   & 1-3 \\
\object{AE~Phe~B}  & F8~V   & $0\fd3624$ & L-M, (P) & B & 4-6 \\
$\sigma^2$~CrB~A   & F9~V   & $1\fd157$  & H,P      & B & 7 \\
\object{HD~307938} & G2~V   & $0\fd5641$ & (L)-P    &   & 8,9 \\
\object{LQ~Lup}    & G2V-IV & $0\fd31$   & (L)-P    &   & 10 \\
\object{HII~3163}  & K0~V   & $0\fd414$  & L, P     &   & 11 \\
\object{AB~Dor}    & K0~V   & $0\fd5148$ & L-P      &   & 12-22 \\
\object{HII~686}   & K4~V   & $0\fd3624$ & L,M,P    &   & 11 \\
\object{LO~Peg}    & K5-7~V & $0\fd4236$ & (L)-P    &   & 23-25 \\
\hline
\end{tabular}
\tablefoot{\tablefoottext{a}{Only Zeeman-Doppler imaging maps, no brightness, spot occupancy, nor temperature maps.}}
\tablebib{
(1)~\citet{2007MNRAS.374L..42C}; (2)~\citet{2008MNRAS.385.1179D}; (3)~\citet{2009MNRAS.398.1383F}; (4)~\citet{1991Msngr..66...47M}; (5)~\citet{1994A&A...288..529M}; (6)~\citet{2004MNRAS.348.1321B}; (7)~\citet{2003A&A...399..315S}; (8)~\citet{2004AN....325..246M}; (9)~\citet{2005MNRAS.359..711M}; (10)~\citet{2000MNRAS.316..699D}; (11)~\citet{1999ApJS..123..251S}; (12)~\citet{1994A&A...289..899K}; (13)~\citet{1994MNRAS.269..814C}; (14)~\citet{1995MNRAS.275..534C}; (15)~\citet{1995MNRAS.277.1145U}; (16)~\citet{1997MNRAS.288..343H}; (17)~\citet{1997MNRAS.290L..37U}; (18)~\citet{1997MNRAS.291....1D}; (19)~\citet{1999MNRAS.302..437D}; (20)~\citet{1999MNRAS.308..493C}; (21)~\citet{2003MNRAS.345.1145D}; (22)~\citet{2007MNRAS.375..567J}; (23)~\citet{1999MNRAS.307..685L}; (24)~\citet{2005MNRAS.356.1501B}; (25)~\citet{2008MNRAS.387..237P}
}
\end{table}

High-latitude spots are thus common features in rapidly rotating dwarfs, and sometimes the maps show the whole polar region to be covered with spots. A more interesting aspect than how these high-latitude/polar spots can be formed would be whether low-latitude spots are detected or not. However, we recall that the maps have been made and visualised using different techniques. Therefore, it is difficult to judge how prominent the low-latitude features are. Furthermore, low-latitude features appear often less resolved as a result of lower spatial resolution. The simulations show that the places of the spots (or spot groups) are correct, but the temperature contrast suffers from the lack of resolution.


\section{ Lithium abundance }

\begin{figure} 
\resizebox{\hsize}{!}{\includegraphics{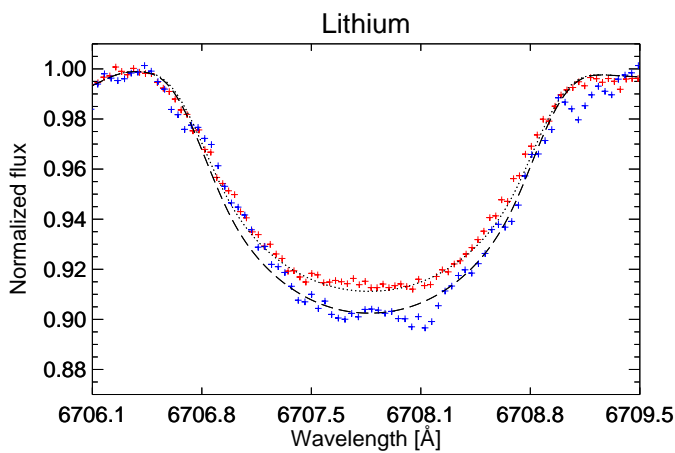}} 
\caption{ Mean lithium-line profiles for the 2008 (red) and 2010 (blue) observations. For comparison, we have plotted synthetic spectra calculated using $\log N{\rm (Li)}=3.25$ (dotted line) and $\log N{\rm (Li)}=3.35$ (dashed line), assuming a fixed temperature in the model calculations.}
\label{lithium}
\end{figure}

\citet{1994A&A...285..272T} have reported a lithium abundance of $\log N{\rm (Li)}=3.2$ for AF~Lep. Our 2005 observations do not cover the lithium line, but for the later observations this region is covered. We determined the lithium-line equivalent width for all individual spectra listed in Tables~\ref{stellaobs} and \ref{aatobs}. To do this, we fitted a double Gaussian to the \ion{Li}{i} 6707.8~\AA\ line and the nearby \ion{Fe}{i} 6707.4~\AA\ line. Although this is still a combined equivalent width of \element[][6]{Li} and \element[][7]{Li}, the Fe blend is effectively removed. Generally, during 2008 the equivalent width of the lithium line was measured to be smaller than in 2009 or 2010. A clear difference is seen in the mean lithium-line profiles (Fig.~\ref{lithium}). As expected (see, e.g., \citealt{1993A&A...267..145P}; \citealt{2002AN....323..325C}), the seasonal equivalent-width variations do not show any clear correlation with the phases of the spots. An excellent review of the relationship between spots and lithium equivalent width variations is given by \citet{1996IAUS..176..345F}. 

To determine the lithium abundance, we calculated synthetic line profiles, including both atomic and CN line blends. The abundance best fitting the 2008 data is $\log N{\rm (Li)}=3.25$, while $\log N{\rm (Li)}=3.35$ best describes data from both 2009 and 2010 (see Fig.~\ref{lithium}) when the temperature is kept constant. The alternative is that the lithium abundance remains constant and the changes detected are due to overall temperature fluctuations. The model grid has temperatures in 250~K steps, and already one step down results in deeper line profiles than any of the observed ones. As Fig.~\ref{lithium} illustrates, the deformation of the lithium line is stronger in 2010 than in 2008. This agrees with the temperature maps, which show that the spot is larger and cooler in 2010 -- although we recall that the phase coverage during 2010 is incomplete. The derived values agree well with the result by \citet{1994A&A...285..272T}, implying that AF~Lep is a young object. Furthermore, the literature gives varying values for the equivalent width ($EW$) of the lithium line on AF~Lep. For example, \citet{2003A&A...399..983W} reported $EW$=191~m\AA, while \citet{2007AJ....133.2524W} reported $EW$=135~m\AA. The values obtained by \citet{2008ApJ...689.1127M} and \citet{2010A&A...517A..88W} are in between, $EW$=146~m\AA\ and $EW$=149~m\AA, respectively. The mean profile of the 2008 observations has an equivalent width of 160~m\AA, while the corresponding value from 2010 mean profile is 182~m\AA.

\section{ Dynamo solutions }

\subsection{ Dynamo setup }

\begin{figure}
\resizebox{0.95\hsize}{!}{\includegraphics{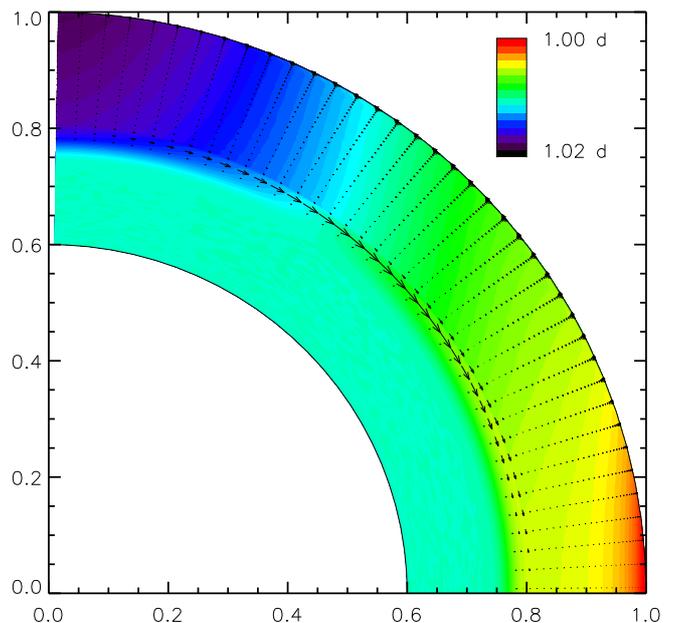}} 
\caption{Rotation profile and meridional circulation from the $\Lambda$-effect model in the vertical cross-section of the computational domain, assuming that the core rotation is the average angular velocity between equator and pole at the bottom of the convection zone. Note that the shear layer near $0.8R_\ast$ is not an artefact but the result of the $\Lambda$-effect model.\label{sps_omega_1}}
\end{figure}

The aim of this section is to use the theoretical results for the differential rotation and meridional circulation computed for a star of $1.2M_\odot$ in a kinematic dynamo model to study whether we can produce a theoretical butterfly diagram that is comparable with the observed spot locations. We chose a spherical shell with $r$, $\theta$, and $\phi$ being the radius, colatitude, and azimuth, and placed the inner boundary at 60\% of the stellar radius $R_\ast$ (which is well within the radiative interior of a late-F star), while the outer radius is the stellar surface. The velocity is a vector field $\vec{u}$, constant in time, containing the theoretical differential rotation and meridional circulation of the star (see, e.g., Fig.~\ref{sps_omega_1}). They were obtained independently from a mean-field hydrodynamic model employing the $\Lambda$-effect and the baroclinicity emerging in rotating, stratified turbulence \citep{1989drsc.book.....R, 1993A&A...279L...1K, 1994A&A...292..125K, 2005AN....326..379K}. A comparison of a large set of differential rotation parameters obtained for low-mass stars from the Kepler mission \citep{2013A&A...560A...4R} and the theoretical predictions by the $\Lambda$-effect theory \citep{2011AN....332..933K} shows a fair agreement and gives us confidence in applying the method to a particular, observed star.

The other effect of rotating, stratified turbulence is the $\alpha$-effect, which gives rise to a turbulent (averaged) electromotive force, generating poloidal magnetic fields from toroidal ones, and vice versa \citep{1980mfmd.book.....K}. The existence of this net effect has been debated recently and may require the removal of small-scale magnetic helicity from the stellar convection zone \citep{2003ApJ...584L..99B, 2011A&A...534A..11W}. We did not consider the exterior here and assumed that it grants the existence of the $\alpha$-effect.

We solved the induction equation in the mean-field formulation with an $\alpha$-effect,
\begin{eqnarray}
  \frac{\partial\vec B}{\partial t} & = &
  \nabla\times\left(\vec u\times\vec B + \vec\alpha(r,\theta,B^2)\circ\vec B -
  \frac{1}{2}(\nabla\eta_{\rm T})\times \vec{B}\right){}
\nonumber\\ 
  & & - \nabla\times\eta_{\rm T} \nabla\times\vec B,
\end{eqnarray}
where $\vec B$ is the large-scale magnetic field and $\vec u$ is a given velocity field. The equation also includes diamagnetic pumping, which is a result of a gradient in the turbulent velocities and often neglected in mean-field dynamos. Whenever a gradient in the turbulent magnetic diffusivity $\eta_{\rm T}$ is implemented, a variation of the turbulence intensity is implied and requires treatment of the diamagnetic pumping \citep{1980mfmd.book.....K,2012SoPh..276....3K}.

\begin{figure}
\resizebox{0.95\hsize}{!}{\includegraphics{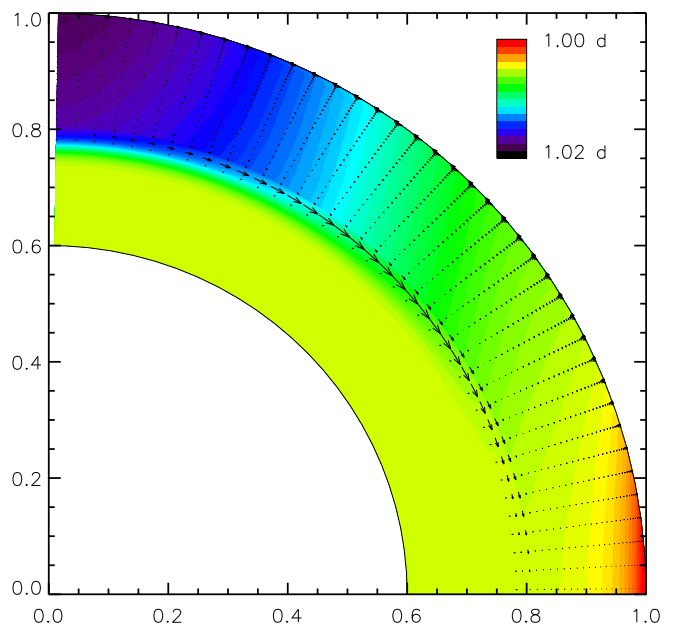}}
\caption{Modified rotation profile with an interior rotating as fast as the bottom of the convection zone at the equator. The meridional circulation is the same as in Fig.~\ref{sps_omega_1}.\label{sps_omega_24}}
\end{figure}

The radius of the star $R_\ast$ and $\eta_{\rm T}$ are used to normalise the induction equation, which then modifies to
\begin{eqnarray}
  \frac{\partial\vec{\hat B}}{\partial t} & = &
  \nabla\times\left(C_\Omega\vec{\hat u}\times\vec{\hat B} + 
  C_\alpha\psi(\hat B^2)\vec{\hat\alpha}(\hat r,\theta)\circ\vec{\hat B} -
  \frac{1}{2}(\nabla\hat\eta_{\rm T})\times \vec{\hat B}\right) {}
\nonumber\\
  & & - \nabla\times \hat\eta_{\rm T} \nabla\times \vec{\hat B},
  \label{normalized}
\end{eqnarray}
where we also split the $\alpha$-tensor into a spatially varying part and 
a part depending on $B^2$, since they are (assumed to be) independent. The resulting dimensionless parameters are $C_\Omega = R_\ast^2\Omega_{\rm eq}/\eta_{\rm T}$ and $C_\alpha = R_\ast \alpha_0/\eta_{\rm T}$, with $\Omega_{\rm eq}$ being the equatorial surface angular velocity and $\alpha_0$ the maximum $\alpha$-effect in the star in m/s.

For better legibility, we omit all hats from the normalised quantities of $r$, $\vec u$, $\vec B$, $\alpha$, and $\eta$ in the following. In general, $\vec{\alpha}$ is a tensor of which our $\alpha$-term contains the symmetric part, while the antisymmetric part is represented by the diamagnetic pumping term, with all three distinct elements of the latter being equal to $-\nabla\eta/2$. The six distinct tensor elements of $\vec{\alpha}$ for fast rotation (anisotropic $\alpha$-effect) are
\begin{eqnarray}
  \alpha_{rr} &=& f(r)\cos\theta\,(1-2\cos^2\theta),\nonumber\\
  \alpha_{\theta\theta}&=& f(r)\cos\theta\,(1-2\sin^2\theta),\nonumber\\
  \alpha_{\phi\phi}&=& f(r)\cos\theta,\nonumber\\
  \alpha_{r\theta} &=& 2 f(r)\cos^2\theta\sin\theta\quad{\rm and}\nonumber\\
  \alpha_{r\phi} &=& \alpha_{\theta\phi} = 0;\nonumber\\
  f(r)&=& \frac{1}{2}\left[1+{\rm erf}\left(\frac{r-r_{\rm cz}}{d}\right)\right],
\label{anisotropy}
\end{eqnarray}
where $r_{\rm cz}=0.8$ is the bottom radius of the convection zone and $d=0.02$ is the thickness of the transition to the radiative interior, motivated by the results for the solar tachocline (e.g., \citealt{2011ApJ...735L..45A}). The convection-zone thickness is based on a stellar model of a $1.2M_\odot$ ZAMS star with solar metallicity, computed with the MESA stellar evolution code \citep{2011ApJS..192....3P}. The quenching function $\psi(B^2)$ represents the suppression of the turbulence by large-scale magnetic fields \citep{1989A&A...213..411B}, that is, the strength of the $\alpha$-effect, and is defined as
\begin{equation}
  \psi(B^2) = \frac{1}{1+B^2/B_{\rm eq}^2}.
\end{equation}
The magnetic field at which the $\alpha$-effect drops significantly is the equipartition field strength, $B_{\rm eq}=\sqrt{\mu_0\rho}\,u_{\rm rms}$, where $\mu_0$ is the permeability constant, $\rho$ the gas density, and $u_{\rm rms}$ the root-mean-square convective velocity.

The magnetic diffusivity in the radiative interior is much lower than the turbulent diffusivity in the convection zone. We employ an arbitrary function for the non-dimensional diffusivity,
\begin{equation}
  \eta_{\rm T} = \eta_{\rm core} + \frac{\eta_{\rm cz}-\eta_{\rm core}}{2}
  \left[1+{\rm erf}\left(\frac{r-r_{\rm cz}}{d}\right)\right],
\end{equation}
where $\eta_{\rm core}=0.01$ is the diffusivity in the interior and $\eta_{\rm cz}=1$ is the turbulent value in the convection zone.

Assuming $R_\ast = 1.1R_\odot$, $P=1.0$~days, and a turbulent diffusivity of $\eta_{\rm T}=10^{12}$~cm$^2$/s, we obtain a magnetic Reynolds number of $C_\Omega=4.09\times10^5$. The meridional flow obtained in conjunction with the differential rotation is $u_{\rm m} = 48$~m/s, giving a flow Reynolds number of 363.1.

The bottom boundary conditions at $r=0.6$ for the magnetic field correspond to perfect conductor conditions, while the top boundary at $r=1$ has exactly vacuum conditions. The mean-field induction Eq.~(\ref{normalized}) is solved with the numerical code by \citet{2000IJNMF..32..773H}. As already indicated, we read the velocity field ${\vec u}=(u_r(r,\theta), u_\theta(r,\theta), u_\phi(r,\theta))$ from the output of the $\Lambda$-effect model for the convection zone. In the latter, the boundary conditions for the flow model were stress-free at the stellar surface and at the bottom of the convection zone. Note that this is not the bottom boundary of the dynamo model. Since the computational domain for the dynamo is larger, the differential rotation was extrapolated to sharply change to a uniform rotation in the radiative interior (as we know it does in the Sun). The choice of the rotation rate of the interior is free in principle, so we chose two cases: one takes the $\Omega$ of mid-latitudes in the convection zone, the other uses the equatorial $\Omega$ of the convection zone. The theoretical rotation patterns $\Omega(r,\theta)$ for AF~Lep are illustrated in Figs.~\ref{sps_omega_1} and~\ref{sps_omega_24}.

\begin{figure*}
\hspace{.4cm}
\resizebox{.45\hsize}{!}{\includegraphics{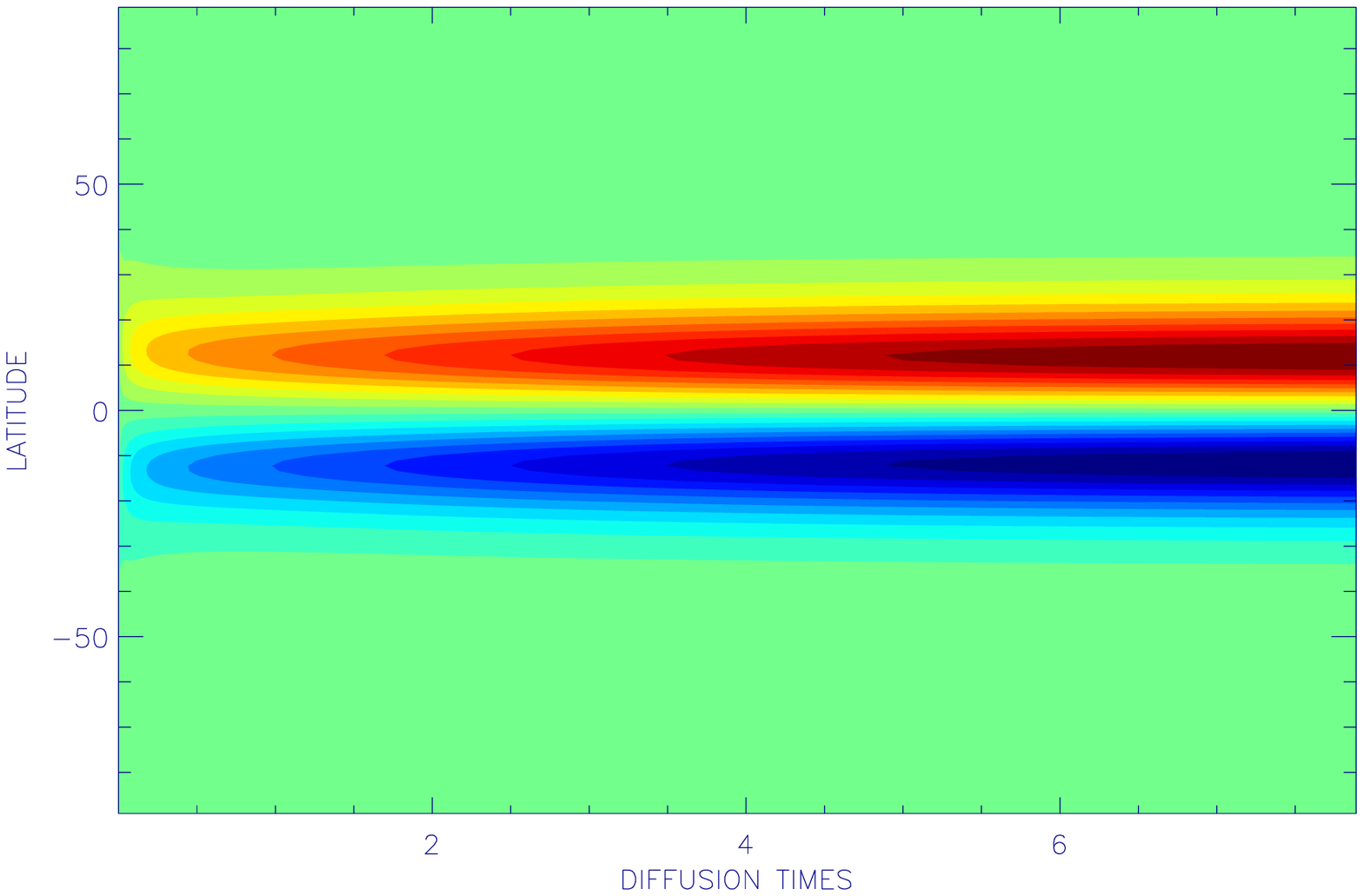}}
\hspace{.5cm}
\resizebox{.45\hsize}{!}{\includegraphics{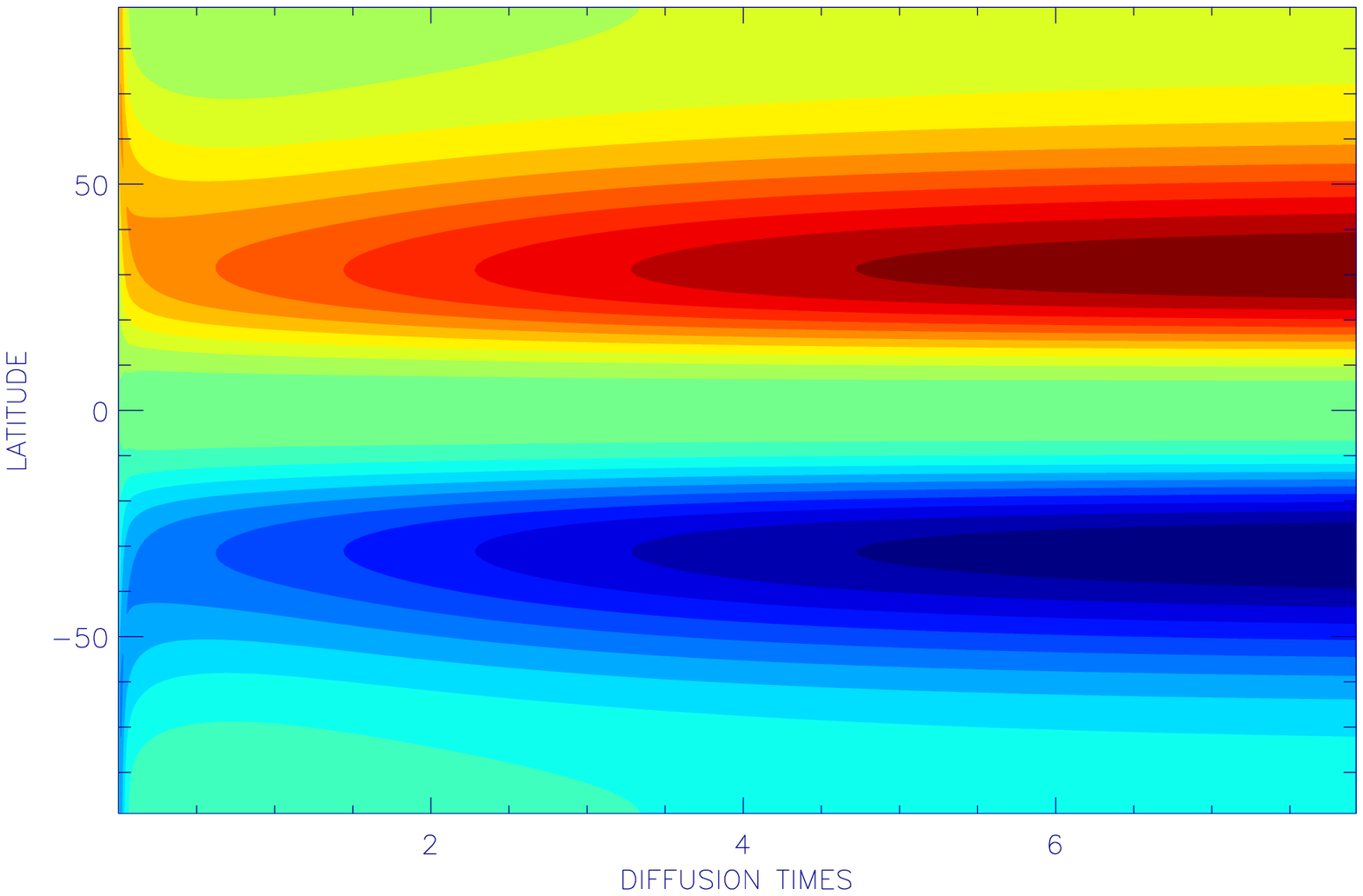}}
\vspace{.3cm}
\caption{Butterfly diagram for the solution with diamagnetic pumping and maximum anisotropy in $\alpha$ according to (\ref{anisotropy}). \emph{Left:} azimuthal magnetic field $B_\phi$ at a radius of $0.8R_\ast$ (maximum value $B_\phi^{\rm max}=0.52$); \emph{right:} radial field $B_r$ at the stellar surface (maximum value $B_r^{\rm max}=0.0007$).
\label{butterfly_aflep_18_08}}
\end{figure*}

\begin{figure*}
\hspace{.4cm}
\resizebox{.45\hsize}{!}{\includegraphics{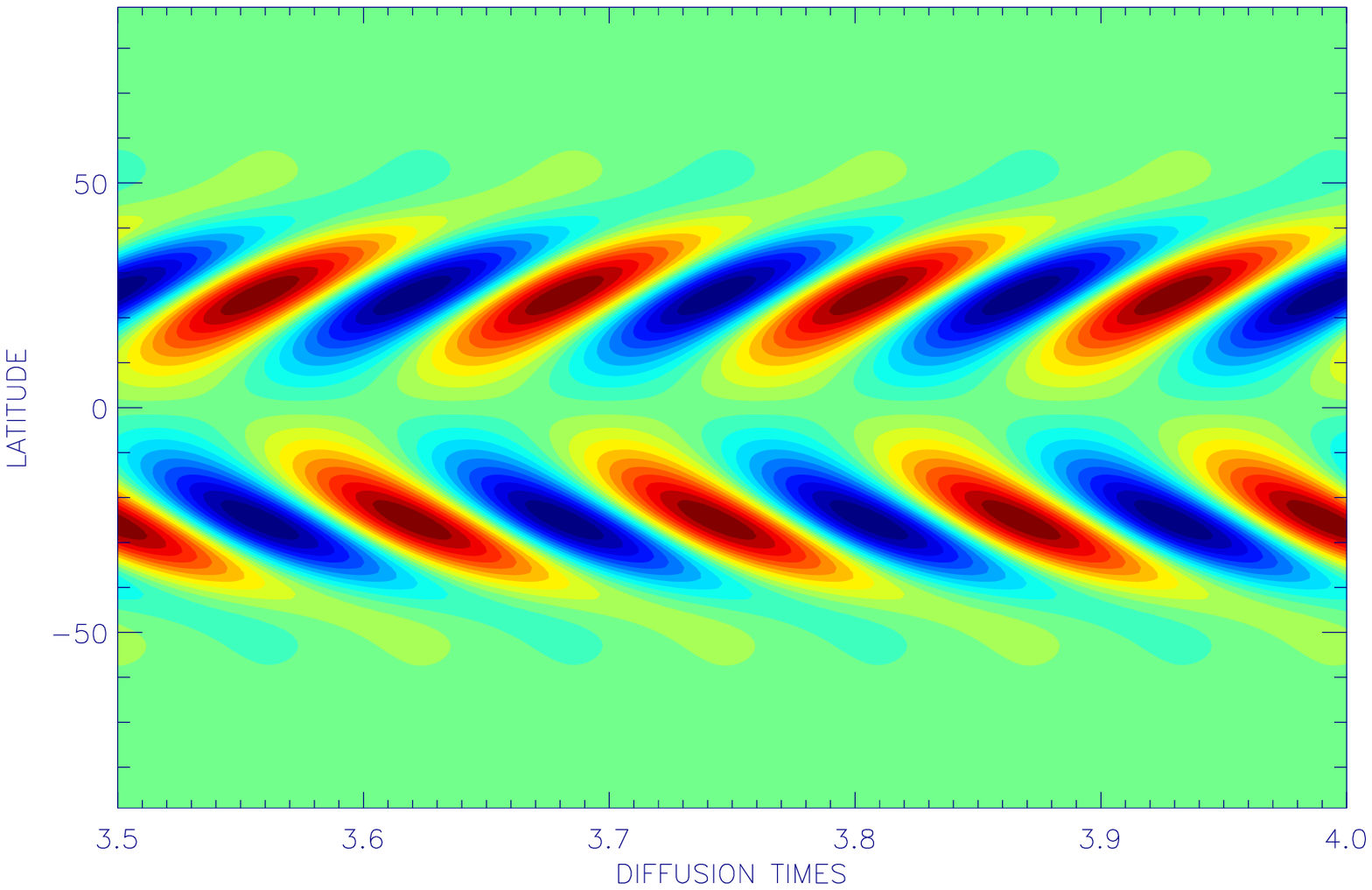}}
\hspace{.5cm}
\resizebox{.45\hsize}{!}{\includegraphics{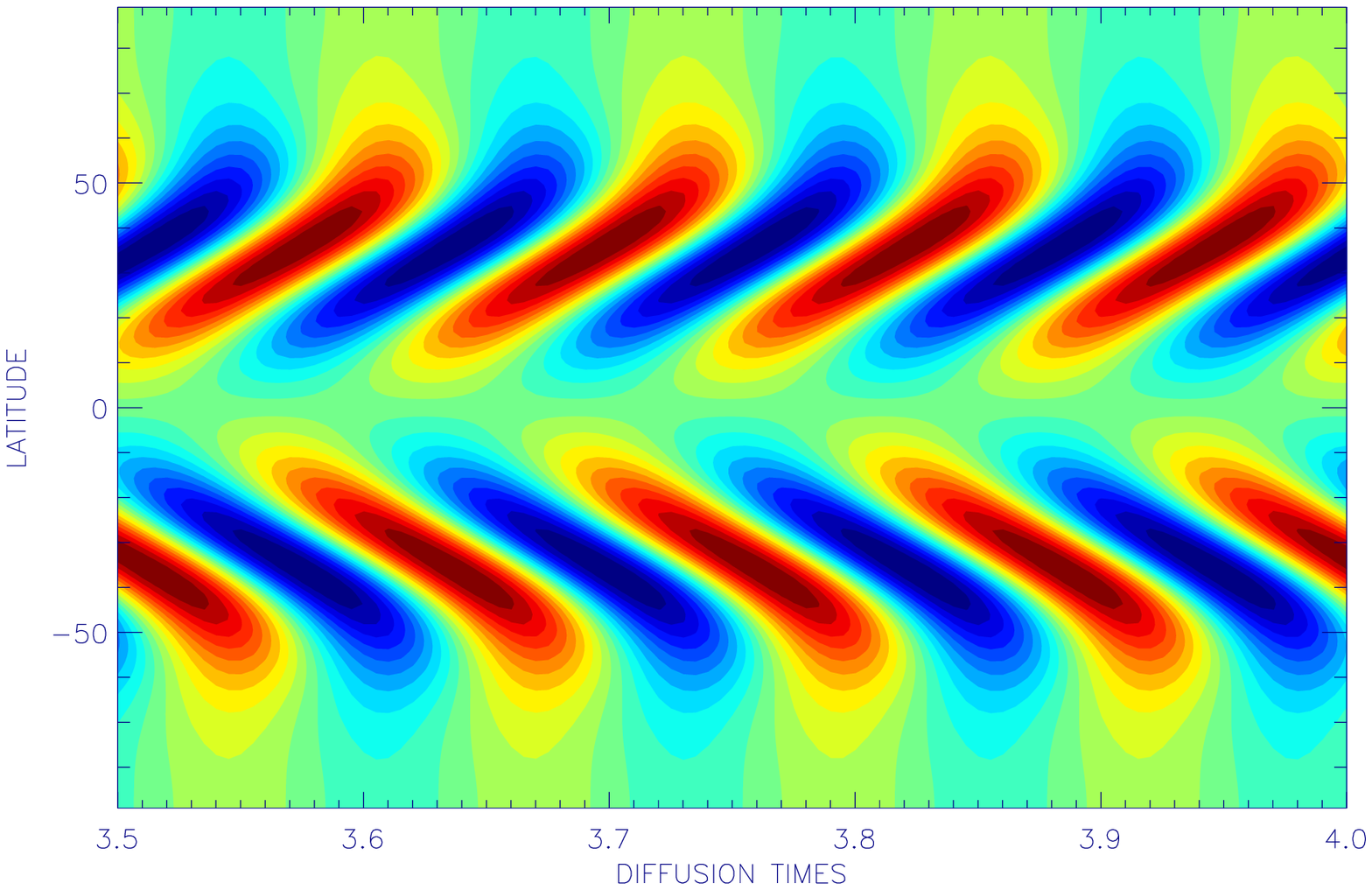}}
\vspace{.3cm}
\caption{As in Fig.~\ref{butterfly_aflep_18_08}, but for a magnetic diffusivity ten times higher. Maximum values are $B_\phi^{\rm max}=0.052$ (left) and $B_r^{\rm max}=0.0095$ (right).
\label{butterfly_aflep_21_08}}
\end{figure*}

\begin{figure*}
\hspace{.4cm}
\resizebox{.45\hsize}{!}{\includegraphics{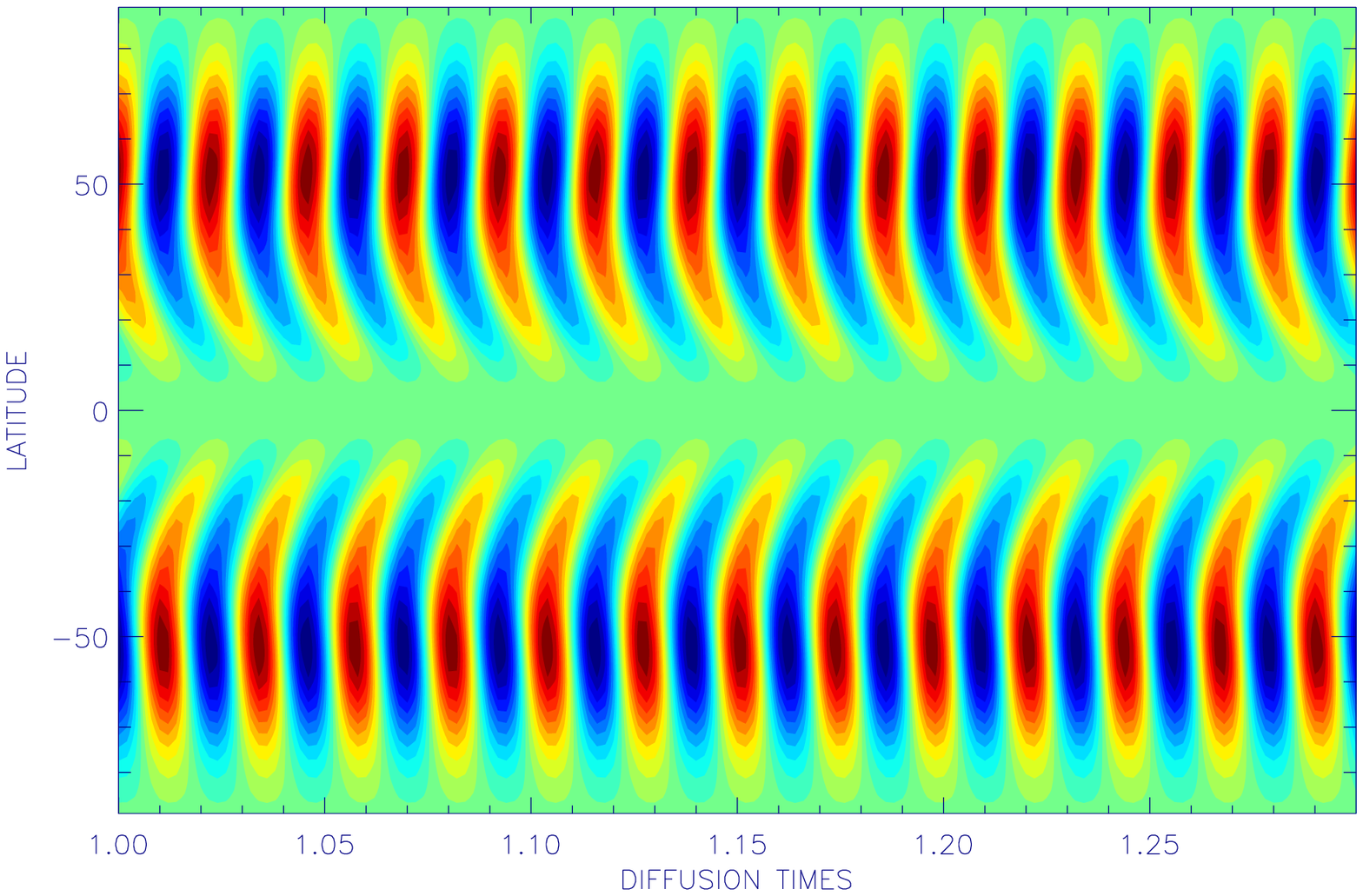}}
\hspace{.5cm}
\resizebox{.45\hsize}{!}{\includegraphics{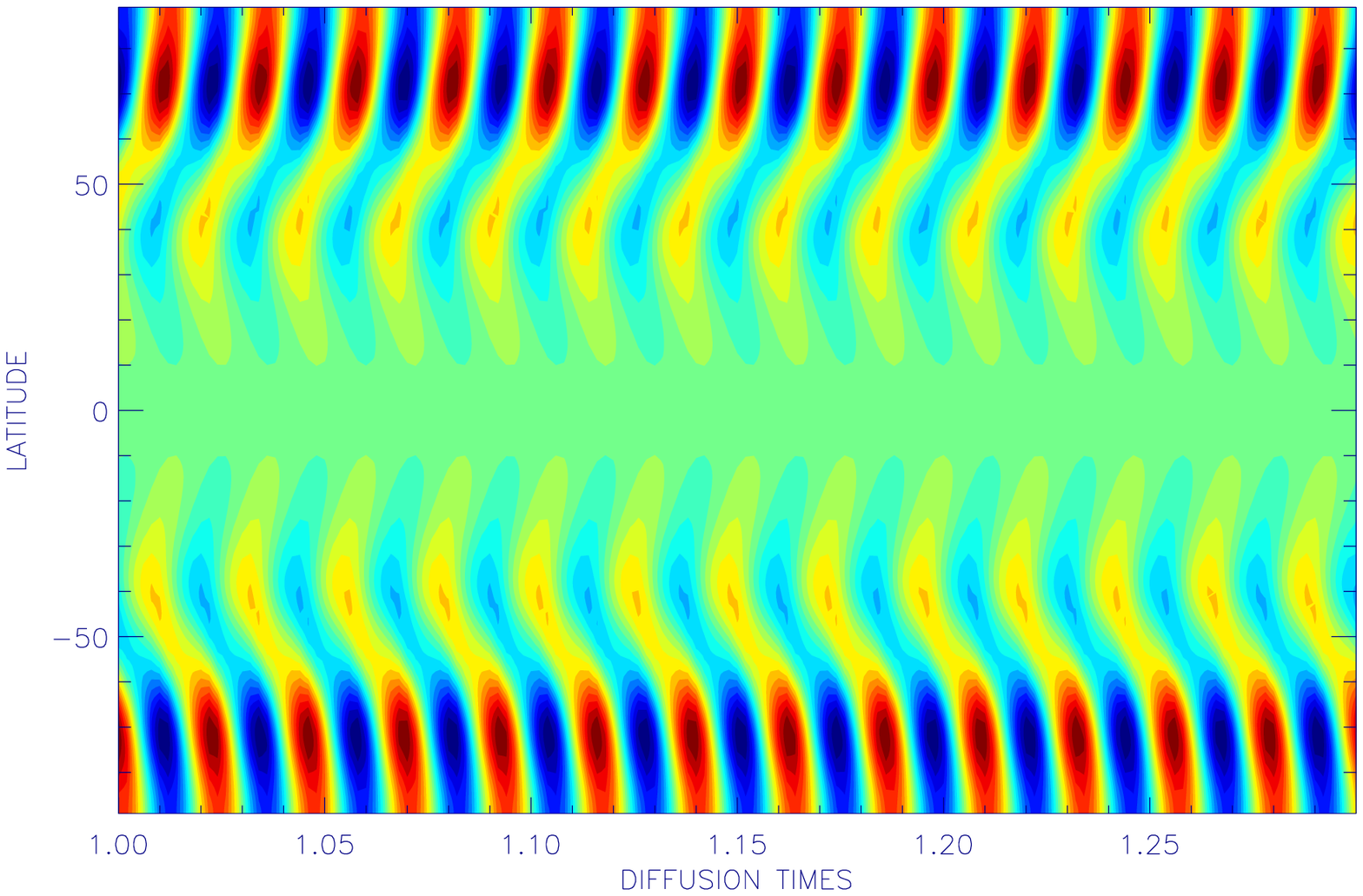}}
\vspace{.3cm}
\caption{As in Fig.~\ref{butterfly_aflep_18_08}, but with a fast-rotating interior (roughly equatorial surface rotation). Maximum values are $B_\phi^{\rm max}=0.76$ (left) and $B_r^{\rm max}=0.044$ (right).
\label{butterfly_aflep_24_08}}
\end{figure*}

\subsection{Results}

The minimum $C_\alpha$ for a growing antisymmetric (dipole-like) solution is 0.20, while it is 1.05 for the symmetric (quadrupole-like) solution. Since magnetic fields suppress the turbulence whence the $\alpha$-effect (quenching), the effective $\alpha$-effect is reduced to its lowest marginal value in any place where fields grow. A symmetric solution is therefore very unlikely in AF~Lep. We concentrate on antisymmetric solutions in the following. For the estimated magnetic Reynolds number of $C_\Omega=4.09\times 10^5$, both antisymetric and symmetric solutions are stationary. Strong meridional circulation can change the oscillatory behaviour of the solutions into stationary ones \citep{2001A&A...374..301K}. 

The latitudinal distribution of the azimuthal magnetic field at the tachocline and the radial field at the surface are plotted over time in Fig.~\ref{butterfly_aflep_18_08}. The plot for the tachocline fields is motivated by the concept that the strong shear at the base of the convection zone generates strong azimuthal fields that eventually become buoyantly unstable and rise to the surface to form spots (not modelled here). The largest tachocline fields reside at low latitudes. With this location, high-emergence latitudes are extremely difficult to achieve, since thin flux-tube simulations do not show tubes emerging at axis distances smaller than the starting distance \citep{2000A&A...355.1087G}. The highest emergence latitude $b_{\rm e}$ for a given initial latitude $b_{\rm i}$ is therefore $\cos b_{\rm e} = r_{\rm cz} \cos b_{\rm i}$. An initial latitude (residence of strong dynamo fields) of $30\degr$ results in a highest emergence latitude of $46\degr$. The largest azimuthal fields of $B_\phi^{\rm max}=5.5$ occur at $r=0.75$ because of the pumping, which is most efficient on stationary fields. The strongest radial surface fields are also constrained to latitudes below about $40\degr$. We do not believe the observed high-latitude fields can be explained with this dynamo solution.

Using an isotropic $\alpha$-effect with $\alpha_{rr} = \alpha_{\theta\theta} = \alpha_{\phi\phi}$ (corresponding to very slow rotation unlike AF~Lep), we also obtain stationary solutions with an unchanged critical $C_\alpha=0.20$. The $\Omega$-effect is just much stronger in the toroidal-field equation than any of the various $\alpha$-components.

The magnetic diffusivity is a relatively poorly known value in the bulk of the convection zones of the Sun and stars. We computed the dynamo solutions for two other values of $\eta_{\rm T}$, $10^{11}$ and $10^{13}$~cm$^2$/s, leading to $C_\Omega=4.09\times10^6$ (LoEta) and $C_\Omega=4.09\times10^4$ (HiEta), respectively. These diffusivities correspond to diffusion times of 1900~yr and 19~yr, again using $R_\ast$ as the length scale. The ratio between rotational and meridional velocity is always kept the same. The LoEta run delivers a stationary solution as well, with a critical $C_\alpha=0.43$. The HiEta run leads to an oscillatory solution with a critical $C_\alpha=10.8$ -- a stronger $\alpha$-effect is now needed to compensate for the weaker $\Omega$-effect. The butterfly diagram shows significant azimuthal fields at the tachocline and radial surface fields only below $60\degr$ latitude (Fig.~\ref{butterfly_aflep_21_08}).

In an attempt to increase the shear at high latitudes, we also employed a modified rotation profile that assumes that the stellar radiative interior rotates with  the same velocity as the bottom of the convection zone {\em at the equator\/} (Fig.~\ref{sps_omega_24}). All other parameters are the same as for Fig.~\ref{butterfly_aflep_18_08}. The solutions are now oscillatory, and the largest azimuthal field resides around $50\degr$ latitude (Fig.~\ref{butterfly_aflep_24_08}). Since fast rotation tends to increase the latitudes of the spot emergence as compared to the tachocline latitude of the flux (e.g., \citealt{2004AN....325..417G}), the spot latitudes from Doppler imaging may not contradict the dynamo solution. The observed spot latitudes can be a tool to probe the internal rotation of low-mass stars. The spin-down of the interior of such a young star may not have finished yet, in contrast to the Sun, where the interior rotates with a velocity seen at mid-latitudes on the surface.

If the value for the magnetic diffusivity is correct, the period of the cycles in Fig.~\ref{butterfly_aflep_24_08} would be 4.5~yr. A possible cycle period of about 4~yr is indicated by the upper panel of Fig.~\ref{photo} (see also Sect.~\ref{sec:phovar}), resulting in a full-cycle period of 8~yr including a field reversal. Given our ignorance of the exact turbulent magnetic diffusivity, the model cycle time is compatible with the observed long-term variability.

 \begin{figure*}
\hspace{.4cm}
\resizebox{.45\hsize}{!}{\includegraphics{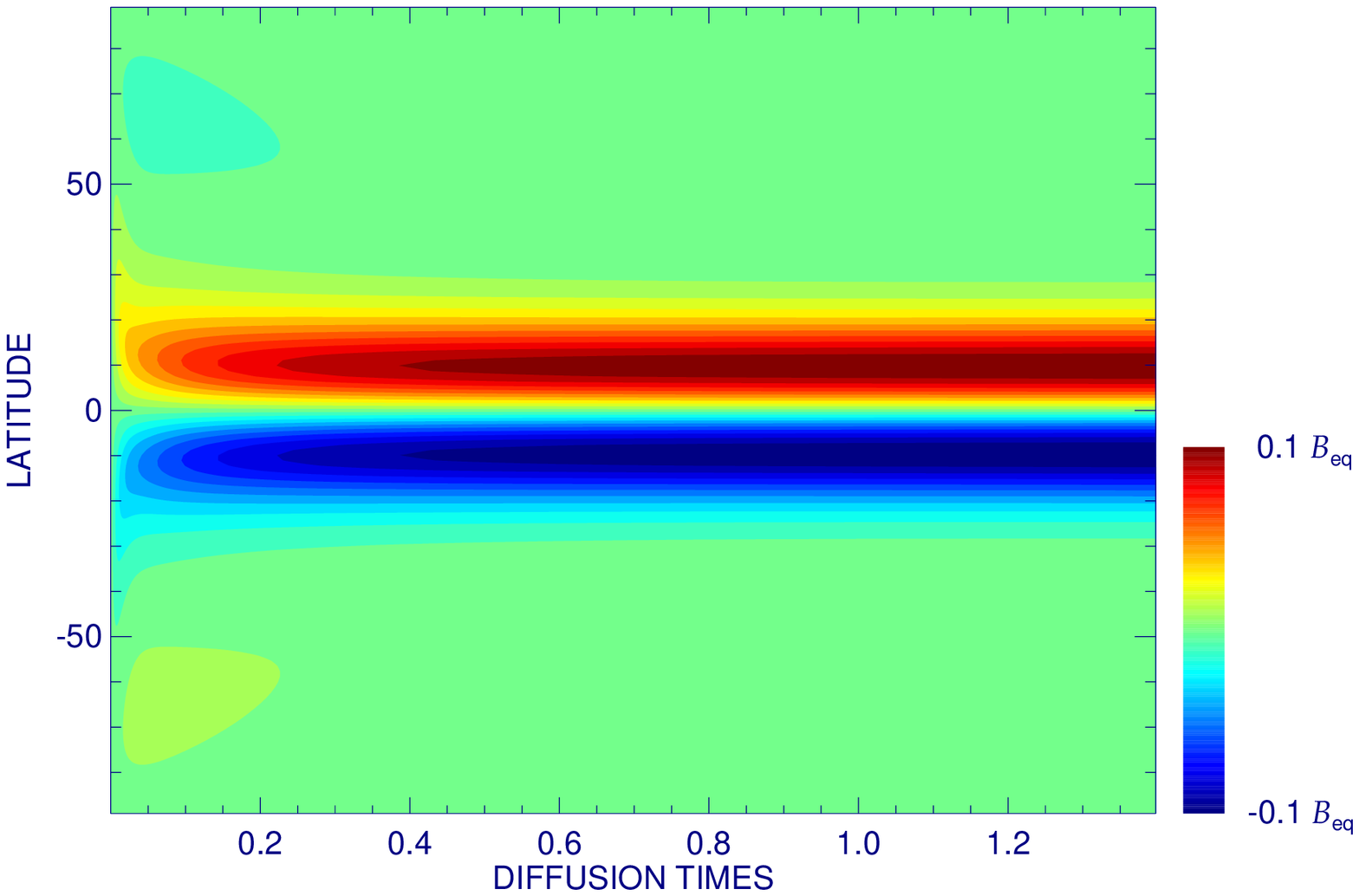}}
\hspace{.5cm}
\resizebox{.45\hsize}{!}{\includegraphics{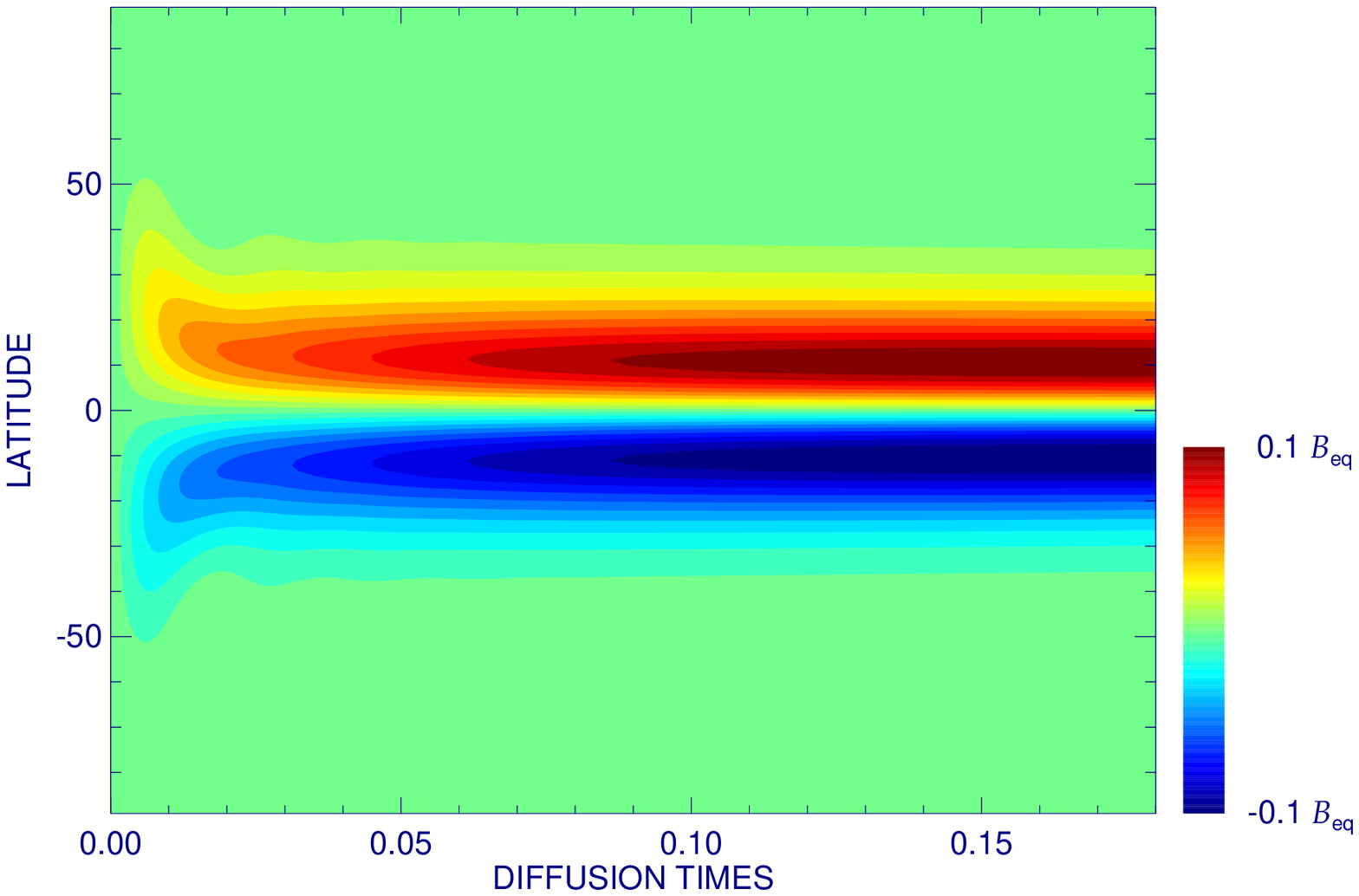}}
\vspace{.3cm}
\caption{Butterfly diagram for the solution with diamagnetic pumping, Babcock-Leighton-type source term for the poloidal magnetic field, and a turbulent magnetic diffusivity $\eta_{\rm T}=10^{12}~{\rm cm}^2{\rm /s}$. {\em Left:\/} classical solar case with a poloidal-field generation by active regions at low latitudes. {\em Right:\/} observationally driven case with a poloidal-field generation around $70\degr$ latitudes. In both cases the toroidal magnetic field at the bottom of the convection zone is plotted because this is the field relevant for spots in the Babcock-Leighton scenario.
\label{butterfly_aflep_22_medeta}}
\end{figure*}

\begin{figure*}
\hspace{.4cm}
\resizebox{.45\hsize}{!}{\includegraphics{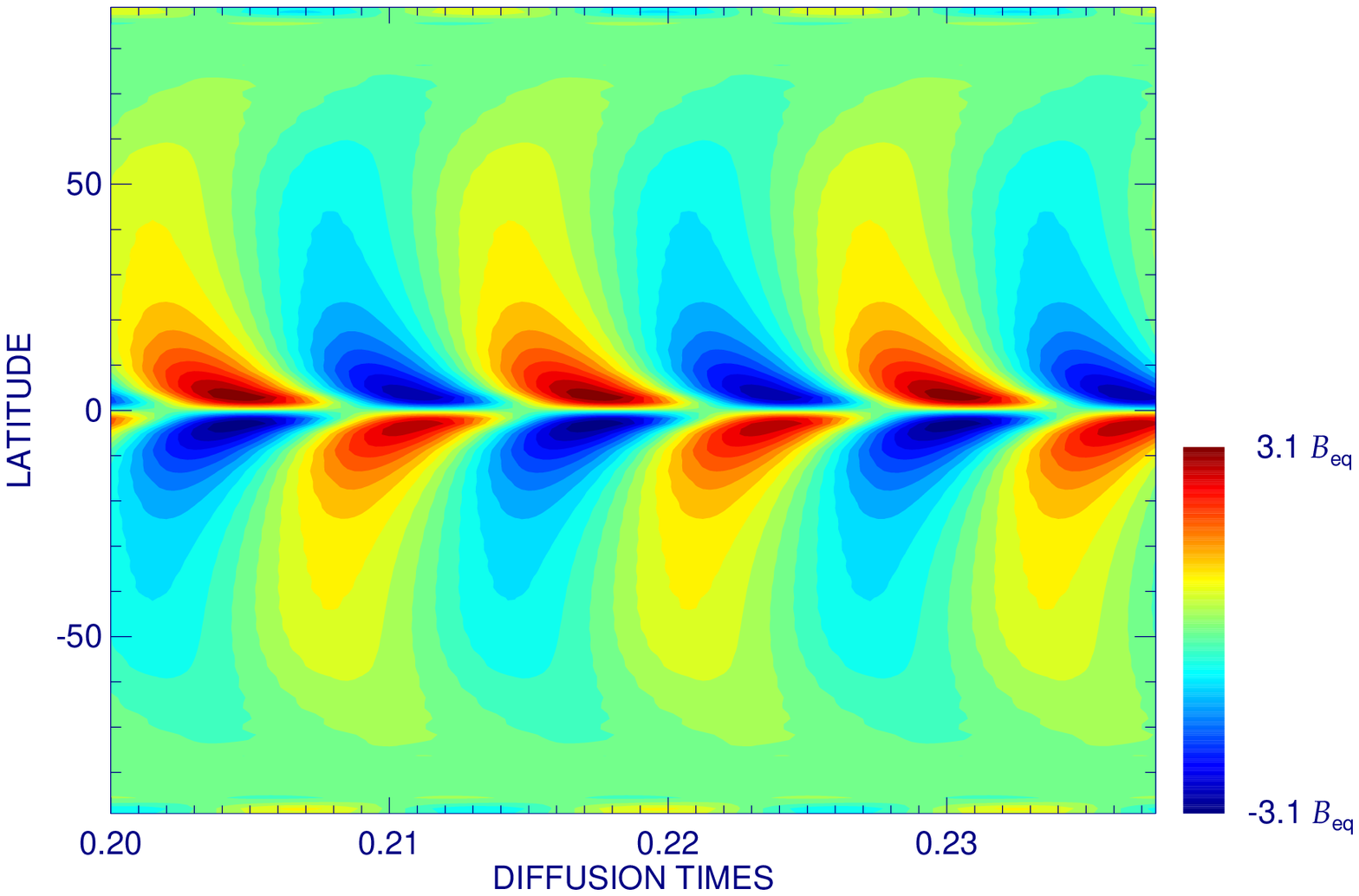}}
\hspace{.5cm}
\resizebox{.45\hsize}{!}{\includegraphics{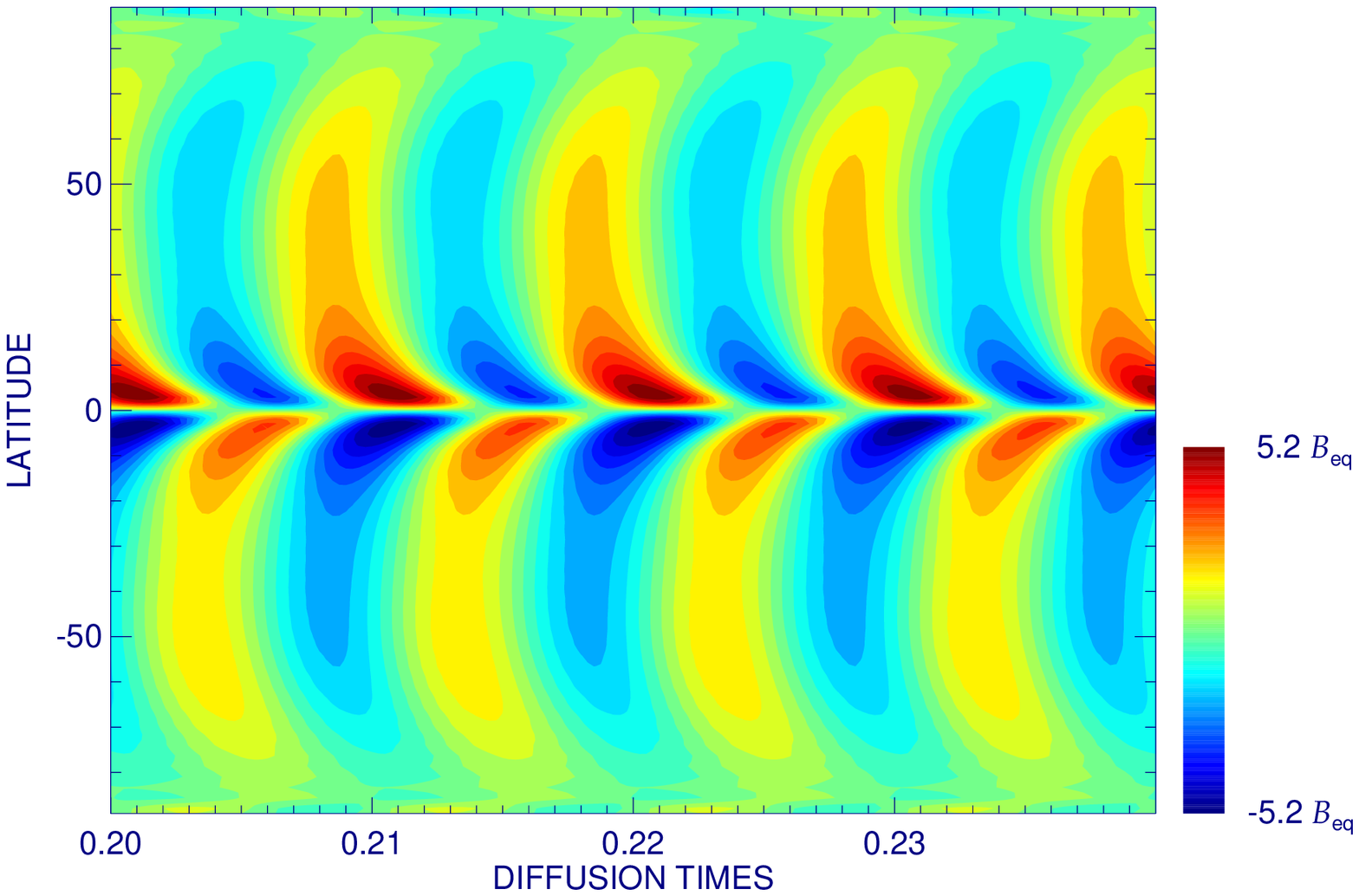}}
\vspace{.3cm}
\caption{Babcock-Leighton-type dynamo solutions with the same setup as in Fig.~\ref{butterfly_aflep_22_medeta}, but for the low turbulent magnetic diffusivity $\eta_{\rm T}=10^{11}~{\rm cm}^2{\rm /s}$. {\em Left:\/} classical solar case with a poloidal-field generation by active regions at low latitudes. {\em Right:\/} observationally driven case with a poloidal-field generation around $70\degr$ latitudes.
\label{butterfly_aflep_22_loweta}}
\end{figure*}

\begin{figure*}
\hspace{.4cm}
\resizebox{.45\hsize}{!}{\includegraphics{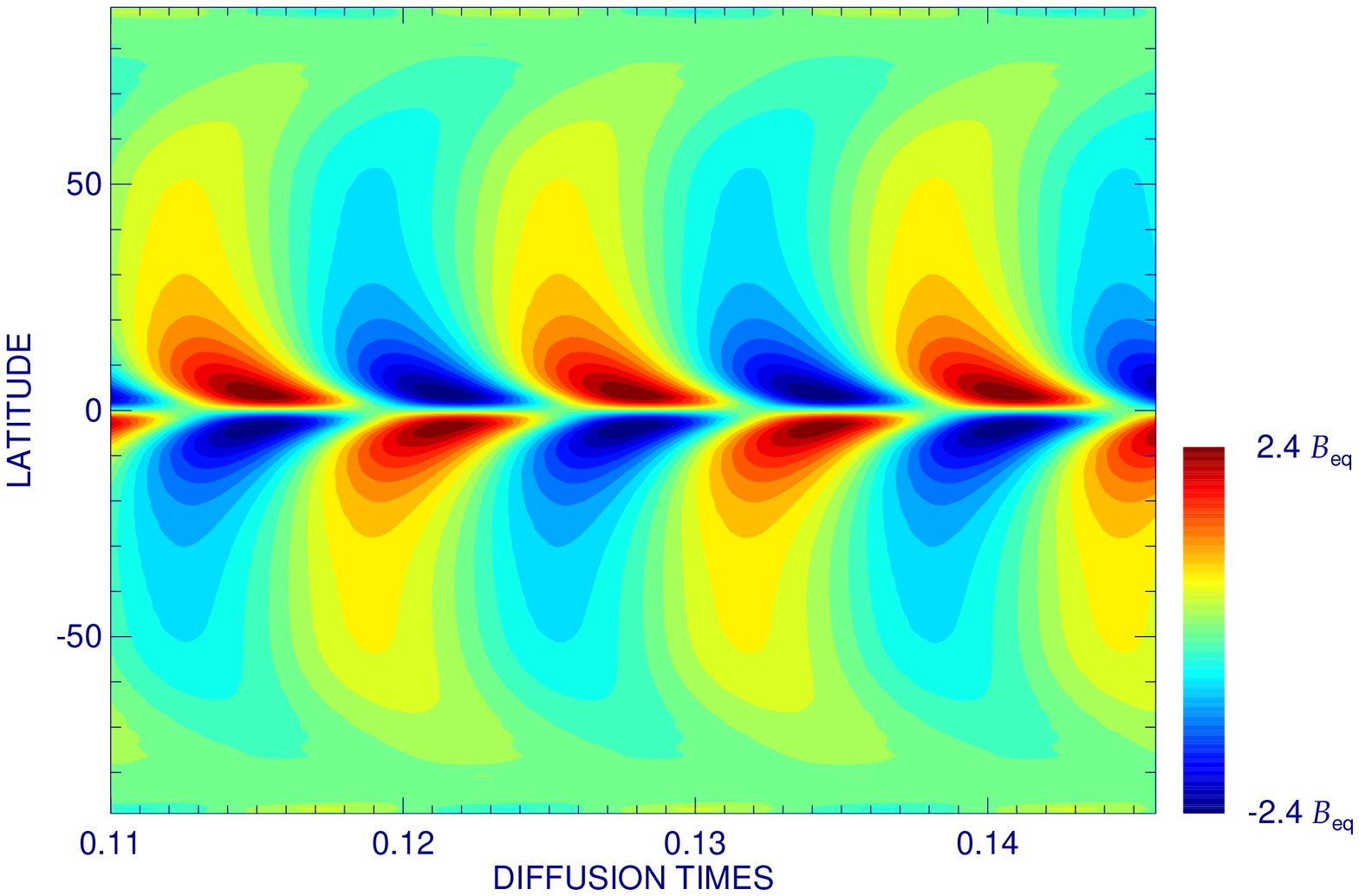}}
\hspace{.5cm}
\resizebox{.45\hsize}{!}{\includegraphics{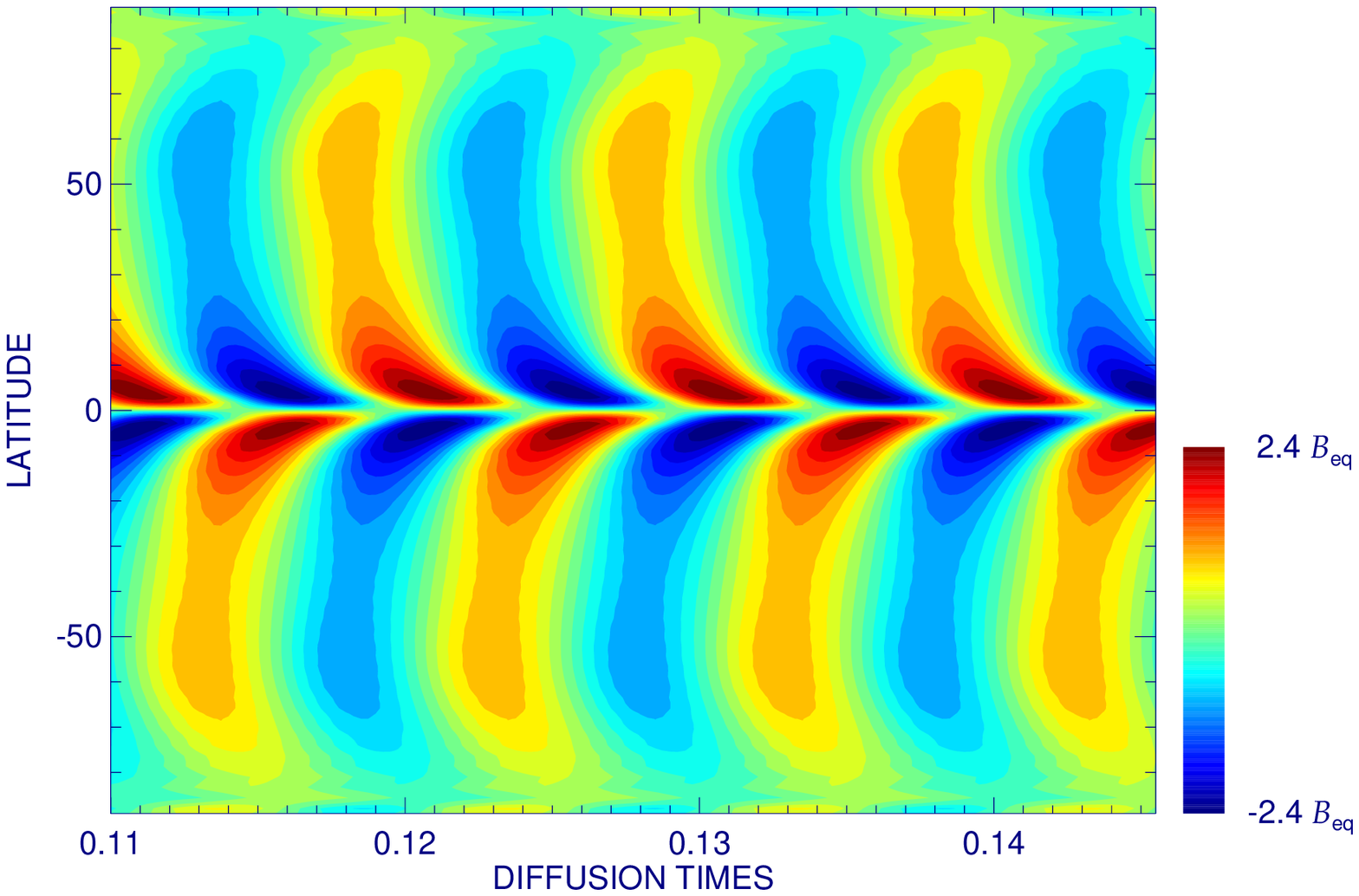}}
\vspace{.3cm}
\caption{Babcock-Leighton-type dynamo solutions with the same setup as in Fig.~\ref{butterfly_aflep_22_loweta}, but for a fast radiation zone. Again, {\em left:\/} poloidal-field generation at low latitudes, and {\em right:\/} poloidal-field generation around $70\degr$ latitudes.
\label{butterfly_aflep_22_fast}}
\end{figure*}

An alternative approach to the solar dynamo is the flux-transport dynamo, which is based on the Babcock-Leighton effect. It has been widely used to explain the features of the solar cycle, but is not based on theoretical consequences of the turbulence. Instead, it assumes that toroidal magnetic flux emerges from the bottom of the convection zone in a time much shorter than the cycle, and it is the tilt of bipolar sunspot groups that contributes to a near-surface poloidal magnetic field. This new poloidal flux is eventually transported down to the bottom of the convection zone by the meridional circulation. The $\alpha$-term in Eq.~\ref{normalized} is now modified to two variants: first, to a distribution typically used in the solar case, and a distribution motivated by the presence of spots on AF~Lep near $70\degr$ latitude. The $\alpha$-term now reads
\begin{equation}
  C_\alpha\psi(B_\phi^2(r_{\rm cz},\theta)){\vec\alpha}(r,\theta)\circ
  {\vec B}(r_{\rm cz},\theta)
\end{equation}
and the $\alpha$-components need to be replaced by $\alpha_{rr} = \alpha_{\theta\theta} = \alpha_{r\theta} = 0$ and
\begin{eqnarray}
  \alpha_{\phi\phi}&=&  \frac{1}{2}\left[1+{\rm erf}
  \left(\frac{r-r_{\rm s}}{d}\right)\right]\cos\theta\sin^2\theta,~{\rm (solar~case)}\nonumber\\
 \alpha_{\phi\phi}&=&  \frac{1}{2}\left[1+{\rm erf}
  \left(\frac{r-r_{\rm s}}{d}\right)\right]\cos\theta\sin^2\left(\,\left|\theta-\frac{\pi}{2}\right|^{2.5}\right),~{\rm (AF~Lep)}\nonumber
\end{eqnarray}
where $r_{\rm s}=0.95$ restricts the poloidal-field production to a near-surface layer. These are purely geometrical descriptions; the only physical requirement is the anti-symmetry about the equator, achieved by $\cos\theta$, as was also used in the third line of Eq.~\ref{anisotropy}. The main change is, however, the non-locality of this ``$\alpha$''-effect because it refers to the azimuthal field at the bottom of the convection zone, but acts at the surface. In contrast to solar flux-transport models, we did not restrict $\alpha_{\phi\phi}$ to low latitudes, since this restriction is motivated by the low emergence latitudes of sunspots, which does not hold here.

When employing the turbulent magnetic diffusivity of $\eta_{\rm T}=10^{12}~{\rm cm}^2{\rm /s}$ as used in the distributed-$\alpha$ dynamos above, we obtain stationary solutions, as shown in Fig.~\ref{butterfly_aflep_22_medeta}. In both cases, the solar-type and the high-latitude Babcock-Leighton effect, the solutions are confined to low latitudes. We omit the surface radial-field plot here, since the mechanism relies on active regions forming as the result of rising flux from the toroidal field at the base of the convection zone.

A second ingredient to a Babcock-Leighton dynamo is a relatively low $\eta_{\rm T}$. Figure~\ref{butterfly_aflep_22_loweta} therefore shows the solutions for a value ten times lower (LoEta), resulting in $C_\Omega=4.09\times10^6$. The solutions are now oscillatory. The solar case shows naturally low-latitude fields, but the high-latitude case also exhibits large toroidal fields at low latitudes, because the meridional circulation, which is equatorward at the base of the convection zone, has a stronger impact on the solutions at this low value of $\eta_{\rm T}$.

Finally, we tested a fast radiative interior as we did for Fig.~\ref{butterfly_aflep_24_08} and repeated the computations of Fig.~\ref{butterfly_aflep_22_loweta}. The results in Fig.~\ref{butterfly_aflep_22_fast} indicate that there is much less influence of the core rotation on the latitude of strongest toroidal fields in the Babcock-Leighton dynamo than in the distributed dynamo. Although there is a medium toroidal field up to latitudes of $60\degr$ in the right panel of Fig.~\ref{butterfly_aflep_22_fast}, the much stronger fields at $<10\degr$ should produce a clear activity pattern not higher than $40\degr$, even for a purely axial flux rise, which is not simulated here.

The various dynamo solutions presented here favour a distributed dynamo in the rapidly rotating convection zone of AF~Lep, which gives results close to the observed spot latitudes if the radiative interior rotates as fast as the near-equator convection zone (unlike the solar radiation zone). We recall that the ZDI maps by \citet{2006ASPC..358..401M} show regions of a near-surface azimuthal field, which also points to a distributed dynamo.


\section{ Summary and discussion}

We have analysed both photometric and spectroscopic observations of the young active F dwarf AF~Lep and studied whether the obtained results can be explained with theoretical models. 

The long-term photometric record of AF~Lep, which covers over 40 years of infrequent observations, shows that the star has become brighter over this time. While the whole period is not sufficiently sampled, the last 4.5 yr of observations are well covered and indicate a cyclic variation. The previously reported rotation period of $\sim$one day implied that finding the true period from ground-based observations is challenging, especially when the photometry also reveals short-term variability due to spot evolution and/or differential rotation. The photometric light curves indicate that the spot coverage can remain stable for months and then undergo several changes from month to month. With time-series analysis methods we have determined the rotation period to be 0.9660 $\pm$ 0.0023 days.

Multi-site spectroscopic observations were used to reconstruct surface temperature maps that all show a dominant high-latitude spot. This indicates that the dynamo operating in this shallow convection zone star must differ from that of the Sun. The mean longitude of the spots drifts with time, but the latitude remains relatively constant (within the accuracy of the temperature maps).

The lithium line of AF~Lep shows some variability over the years. Because of the resolution (250~K) of the model calculations we cannot rule out that the variations are simply due to overall temperature fluctuations. A fluctuating Li abundance seems quite unlikely. Our results, as well as the results from others, have shown that the equivalent width of the lithium line keeps changing. However, the seasonal variations of individual equivalent-width measurements do not show any clear correlation with the phases of the spots.

We considered various dynamo solutions to produce a theoretical butterfly diagram that is comparable with the observed spot locations. A dynamo model based on a differential rotation profile in which the stellar interior rotates at the average surface angular velocity is not able to reproduce the high-latitude spots observed on AF~Lep and also on other F dwarfs. However, if the stellar interior rotates with the same speed as the convection zone at the equator, it is possible to generate a substantial magnetic field at high latitudes of about $60\degr$. Such fast internal rotation may be compatible with the young age of AF~Lep.

It is rather common that ZDI maps of the stars show the magnetic flux spread over a wide range of latitudes. Over the years, the maps have started to show more and more complex field topologies, and increasingly weaker field strengths are detected as well. \citet{2013IAUS..294..447S} compared independent magnetic maps made by two codes from almost simultaneous observations (see \citealt{2010MNRAS.403..159S} and \citealt{2012A&A...548A..95C}). The disparity of the two reconstructions is noticeable. As stated by \citet{2013IAUS..294..447S}, one can only speculate about the reasons for the differences. Clearly more comparisons of the codes are needed and all reconstructions have to be interpreted with care.


\begin{acknowledgements}
This work has made use of the VALD database, operated at Uppsala University, the Institute of Astronomy RAS in Moscow, and the University of Vienna. The authors thank the anonymous referee for useful and enlightening comments and suggestions.
\end{acknowledgements}


\bibliographystyle{aa}
\bibliography{aa24229-14}

\begin{thebibliography}{99}
\expandafter\ifx\csname natexlab\endcsname\relax\def\natexlab#1{#1}\fi

\bibitem[{{Antia} \& {Basu}(2011)}]{2011ApJ...735L..45A}
{Antia}, H.~M. \& {Basu}, S. 2011, \apjl, 735, L45

\bibitem[{{Barnes} {et~al.}(2005{\natexlab{a}}){Barnes}, {Collier Cameron},
  {Donati}, {James}, {Marsden}, \& {Petit}}]{2005MNRAS.357L...1B}
{Barnes}, J.~R., {Collier Cameron}, A., {Donati}, J.-F., {et~al.}
  2005{\natexlab{a}}, \mnras, 357, L1

\bibitem[{{Barnes} {et~al.}(2005{\natexlab{b}}){Barnes}, {Collier Cameron},
  {Lister}, {Pointer}, \& {Still}}]{2005MNRAS.356.1501B}
{Barnes}, J.~R., {Collier Cameron}, A., {Lister}, T.~A., {Pointer}, G.~R., \&
  {Still}, M.~D. 2005{\natexlab{b}}, \mnras, 356, 1501

\bibitem[{{Barnes} {et~al.}(2004){Barnes}, {Lister}, {Hilditch}, \& {Collier
  Cameron}}]{2004MNRAS.348.1321B}
{Barnes}, J.~R., {Lister}, T.~A., {Hilditch}, R.~W., \& {Collier Cameron}, A.
  2004, \mnras, 348, 1321

\bibitem[{{Berdyugina}(1991)}]{1991BCrAO..83...89B}
{Berdyugina}, S.~V. 1991, Bulletin Crimean Astrophysical Observatory, 83, 89

\bibitem[{{Berdyugina}(1998)}]{1998A&A...338...97B}
{Berdyugina}, S.~V. 1998, \aap, 338, 97

\bibitem[{{Berdyugina} {et~al.}(2002){Berdyugina}, {Pelt}, \&
  {Tuominen}}]{2002A&A...394..505B}
{Berdyugina}, S.~V., {Pelt}, J., \& {Tuominen}, I. 2002, \aap, 394, 505

\bibitem[{{Binks} \& {Jeffries}(2014)}]{2014MNRAS.438L..11B}
{Binks}, A.~S. \& {Jeffries}, R.~D. 2014, \mnras, 438, L11

\bibitem[{{Blackman} \& {Brandenburg}(2003)}]{2003ApJ...584L..99B}
{Blackman}, E.~G. \& {Brandenburg}, A. 2003, \apjl, 584, L99

\bibitem[{{Blackwell} \& {Lynas-Gray}(1994)}]{1994A&A...282..899B}
{Blackwell}, D.~E. \& {Lynas-Gray}, A.~E. 1994, \aap, 282, 899

\bibitem[{{Brandenburg} {et~al.}(1989){Brandenburg}, {Krause}, {Meinel},
  {Moss}, \& {Tuominen}}]{1989A&A...213..411B}
{Brandenburg}, A., {Krause}, F., {Meinel}, R., {Moss}, D., \& {Tuominen}, I.
  1989, \aap, 213, 411

\bibitem[{{Budding} {et~al.}(2002){Budding}, {Carter}, {Mengel}, {Slee}, \&
  {Donati}}]{2002PASA...19..527B}
{Budding}, E., {Carter}, B.~D., {Mengel}, M.~W., {Slee}, O.~B., \& {Donati},
  J.-F. 2002, \pasa, 19, 527

\bibitem[{{Budding} {et~al.}(2003){Budding}, {Heckert}, {Soydugan}, {Soydugan},
  {Bakis}, {Bakis}, \& {Erdem}}]{2003IBVS.5451....1B}
{Budding}, E., {Heckert}, P., {Soydugan}, F., {et~al.} 2003, Information
  Bulletin on Variable Stars, 5451, 1

\bibitem[{{Carroll} {et~al.}(2012){Carroll}, {Strassmeier}, {Rice}, \&
  {K{\"u}nstler}}]{2012A&A...548A..95C}
{Carroll}, T.~A., {Strassmeier}, K.~G., {Rice}, J.~B., \& {K{\"u}nstler}, A.
  2012, \aap, 548, A95

\bibitem[{{Catala} {et~al.}(2007){Catala}, {Donati}, {Shkolnik}, {Bohlender},
  \& {Alecian}}]{2007MNRAS.374L..42C}
{Catala}, C., {Donati}, J.-F., {Shkolnik}, E., {Bohlender}, D., \& {Alecian},
  E. 2007, \mnras, 374, L42

\bibitem[{{Cole} {et~al.}(2014){Cole}, {K{\"a}pyl{\"a}}, {Mantere}, \&
  {Brandenburg}}]{2014ApJ...780L..22C}
{Cole}, E., {K{\"a}pyl{\"a}}, P.~J., {Mantere}, M.~J., \& {Brandenburg}, A.
  2014, \apjl, 780, L22

\bibitem[{{Collier Cameron}(1995)}]{1995MNRAS.275..534C}
{Collier Cameron}, A. 1995, \mnras, 275, 534

\bibitem[{{Collier-Cameron} \& {Unruh}(1994)}]{1994MNRAS.269..814C}
{Collier-Cameron}, A. \& {Unruh}, Y.~C. 1994, \mnras, 269, 814

\bibitem[{{Collier Cameron} {et~al.}(1999){Collier Cameron}, {Walter}, {Vilhu},
  {B{\"o}hm}, {Catala}, {Char}, {Clarke}, {Felenbok}, {Foing}, {Ghosh}, {Hao},
  {Huang}, {Jackson}, {Janot-Pacheco}, {Jiang}, {Lagrange}, {Suntzeff}, \&
  {Zhai}}]{1999MNRAS.308..493C}
{Collier Cameron}, A., {Walter}, F.~M., {Vilhu}, O., {et~al.} 1999, \mnras,
  308, 493

\bibitem[{{Cutispoto}(2002)}]{2002AN....323..325C}
{Cutispoto}, G. 2002, Astronomische Nachrichten, 323, 325

\bibitem[{{Cutispoto} {et~al.}(1991){Cutispoto}, {Tagliaferri}, {Giommi},
  {Gouiffes}, {Pallavicini}, {Pasquini}, \& {Rodono}}]{1991A&AS...87..233C}
{Cutispoto}, G., {Tagliaferri}, G., {Giommi}, P., {et~al.} 1991, \aaps, 87, 233

\bibitem[{{Cutispoto} {et~al.}(1996){Cutispoto}, {Tagliaferri}, {Pallavicini},
  {Pasquini}, \& {Rodono}}]{1996A&AS..115...41C}
{Cutispoto}, G., {Tagliaferri}, G., {Pallavicini}, R., {Pasquini}, L., \&
  {Rodono}, M. 1996, \aaps, 115, 41

\bibitem[{{Donati} \& {Collier Cameron}(1997)}]{1997MNRAS.291....1D}
{Donati}, J.-F. \& {Collier Cameron}, A. 1997, \mnras, 291, 1

\bibitem[{{Donati} {et~al.}(1999){Donati}, {Collier Cameron}, {Hussain}, \&
  {Semel}}]{1999MNRAS.302..437D}
{Donati}, J.-F., {Collier Cameron}, A., {Hussain}, G.~A.~J., \& {Semel}, M.
  1999, \mnras, 302, 437

\bibitem[{{Donati} {et~al.}(2003){Donati}, {Collier Cameron}, {Semel},
  {Hussain}, {Petit}, {Carter}, {Marsden}, {Mengel}, {L{\'o}pez Ariste},
  {Jeffers}, \& {Rees}}]{2003MNRAS.345.1145D}
{Donati}, J.-F., {Collier Cameron}, A., {Semel}, M., {et~al.} 2003, \mnras,
  345, 1145

\bibitem[{{Donati} {et~al.}(2000){Donati}, {Mengel}, {Carter}, {Marsden},
  {Collier Cameron}, \& {Wichmann}}]{2000MNRAS.316..699D}
{Donati}, J.-F., {Mengel}, M., {Carter}, B.~D., {et~al.} 2000, \mnras, 316, 699

\bibitem[{{Donati} {et~al.}(2008){Donati}, {Moutou}, {Far{\`e}s}, {Bohlender},
  {Catala}, {Deleuil}, {Shkolnik}, {Collier Cameron}, {Jardine}, \&
  {Walker}}]{2008MNRAS.385.1179D}
{Donati}, J.-F., {Moutou}, C., {Far{\`e}s}, R., {et~al.} 2008, \mnras, 385,
  1179

\bibitem[{{Donati} {et~al.}(1997){Donati}, {Semel}, {Carter}, {Rees}, \&
  {Collier Cameron}}]{1997MNRAS.291..658D}
{Donati}, J.-F., {Semel}, M., {Carter}, B.~D., {Rees}, D.~E., \& {Collier
  Cameron}, A. 1997, \mnras, 291, 658

\bibitem[{{Eggen}(1986)}]{1986AJ.....92..910E}
{Eggen}, O.~J. 1986, \aj, 92, 910

\bibitem[{{Fares} {et~al.}(2009){Fares}, {Donati}, {Moutou}, {Bohlender},
  {Catala}, {Deleuil}, {Shkolnik}, {Collier Cameron}, {Jardine}, \&
  {Walker}}]{2009MNRAS.398.1383F}
{Fares}, R., {Donati}, J.-F., {Moutou}, C., {et~al.} 2009, \mnras, 398, 1383

\bibitem[{{Fekel}(1996)}]{1996IAUS..176..345F}
{Fekel}, F.~C. 1996, in IAU Symposium, Vol. 176, Stellar Surface Structure, ed.
  K.~G. {Strassmeier} \& J.~L. {Linsky}, 345

\bibitem[{{Fern{\'a}ndez} {et~al.}(2008){Fern{\'a}ndez}, {Figueras}, \&
  {Torra}}]{2008A&A...480..735F}
{Fern{\'a}ndez}, D., {Figueras}, F., \& {Torra}, J. 2008, \aap, 480, 735

\bibitem[{{Granzer}(2004)}]{2004AN....325..417G}
{Granzer}, T. 2004, Astronomische Nachrichten, 325, 417

\bibitem[{{Granzer} {et~al.}(2001{\natexlab{a}}){Granzer}, {Reegen}, \&
  {Strassmeier}}]{2001AN....322..325G}
{Granzer}, T., {Reegen}, P., \& {Strassmeier}, K.~G. 2001{\natexlab{a}},
  Astronomische Nachrichten, 322, 325

\bibitem[{{Granzer} {et~al.}(2000){Granzer}, {Sch{\"u}ssler}, {Caligari}, \&
  {Strassmeier}}]{2000A&A...355.1087G}
{Granzer}, T., {Sch{\"u}ssler}, M., {Caligari}, P., \& {Strassmeier}, K.~G.
  2000, \aap, 355, 1087

\bibitem[{{Granzer} {et~al.}(2001{\natexlab{b}}){Granzer}, {Weber}, \&
  {Strassmeier}}]{2001AN....322..295G}
{Granzer}, T., {Weber}, M., \& {Strassmeier}, K.~G. 2001{\natexlab{b}},
  Astronomische Nachrichten, 322, 295

\bibitem[{{Hollerbach}(2000)}]{2000IJNMF..32..773H}
{Hollerbach}, R. 2000, International Journal for Numerical Methods in Fluids,
  32, 773

\bibitem[{{Hussain}(2002)}]{2002AN....323..349H}
{Hussain}, G.~A.~J. 2002, Astronomische Nachrichten, 323, 349

\bibitem[{{Hussain} {et~al.}(1997){Hussain}, {Unruh}, \& {Collier
  Cameron}}]{1997MNRAS.288..343H}
{Hussain}, G.~A.~J., {Unruh}, Y.~C., \& {Collier Cameron}, A. 1997, \mnras,
  288, 343

\bibitem[{{J{\"a}rvinen} \& {Berdyugina}(2010)}]{2010A&A...521A..86J}
{J{\"a}rvinen}, S.~P. \& {Berdyugina}, S.~V. 2010, \aap, 521, A86

\bibitem[{{Jeffers} {et~al.}(2007){Jeffers}, {Donati}, \& {Collier
  Cameron}}]{2007MNRAS.375..567J}
{Jeffers}, S.~V., {Donati}, J.-F., \& {Collier Cameron}, A. 2007, \mnras, 375,
  567

\bibitem[{{Jetsu} \& {Pelt}(1999)}]{1999A&AS..139..629J}
{Jetsu}, L. \& {Pelt}, J. 1999, \aaps, 139, 629

\bibitem[{{Kim} \& {Demarque}(1996)}]{1996ApJ...457..340K}
{Kim}, Y.-C. \& {Demarque}, P. 1996, \apj, 457, 340

\bibitem[{{Kitchatinov} \& {Olemskoy}(2012)}]{2012SoPh..276....3K}
{Kitchatinov}, L.~L. \& {Olemskoy}, S.~V. 2012, \solphys, 276, 3

\bibitem[{{Kitchatinov} \& {R{\"u}diger}(2005)}]{2005AN....326..379K}
{Kitchatinov}, L.~L. \& {R{\"u}diger}, G. 2005, Astronomische Nachrichten, 326,
  379

\bibitem[{{Kitchatinov} {et~al.}(1994){Kitchatinov}, {Ruediger}, \&
  {Kueker}}]{1994A&A...292..125K}
{Kitchatinov}, L.~L., {Ruediger}, G., \& {Kueker}, M. 1994, \aap, 292, 125

\bibitem[{{Krause} \& {Raedler}(1980)}]{1980mfmd.book.....K}
{Krause}, F. \& {Raedler}, K.~H. 1980, {Mean-field magnetohydrodynamics and
  dynamo theory}

\bibitem[{{Kueker} {et~al.}(1993){Kueker}, {Ruediger}, \&
  {Kitchatinov}}]{1993A&A...279L...1K}
{Kueker}, M., {Ruediger}, G., \& {Kitchatinov}, L.~L. 1993, \aap, 279, L1

\bibitem[{{Kuerster} {et~al.}(1994){Kuerster}, {Schmitt}, \&
  {Cutispoto}}]{1994A&A...289..899K}
{Kuerster}, M., {Schmitt}, J.~H.~M.~M., \& {Cutispoto}, G. 1994, \aap, 289, 899

\bibitem[{{K{\"u}ker} \& {R{\"u}diger}(2007)}]{2007AN....328.1050K}
{K{\"u}ker}, M. \& {R{\"u}diger}, G. 2007, Astronomische Nachrichten, 328, 1050

\bibitem[{{K{\"u}ker} \& {R{\"u}diger}(2011)}]{2011AN....332..933K}
{K{\"u}ker}, M. \& {R{\"u}diger}, G. 2011, Astronomische Nachrichten, 332, 933

\bibitem[{{K{\"u}ker} {et~al.}(2001){K{\"u}ker}, {R{\"u}diger}, \&
  {Schultz}}]{2001A&A...374..301K}
{K{\"u}ker}, M., {R{\"u}diger}, G., \& {Schultz}, M. 2001, \aap, 374, 301

\bibitem[{{Kupka} {et~al.}(1999){Kupka}, {Piskunov}, {Ryabchikova}, {Stempels},
  \& {Weiss}}]{1999A&AS..138..119K}
{Kupka}, F., {Piskunov}, N., {Ryabchikova}, T.~A., {Stempels}, H.~C., \&
  {Weiss}, W.~W. 1999, \aaps, 138, 119

\bibitem[{{Kurucz}(1993)}]{1993KurCD..13.....K}
{Kurucz}, R. 1993, ATLAS9 Stellar Atmosphere Programs and 2 km/s grid.~Kurucz
  CD-ROM No.~13.~ Cambridge, Mass.: Smithsonian Astrophysical Observatory,
  1993., 13

\bibitem[{{Lanza}(2011)}]{2011IAUS..273...89L}
{Lanza}, A.~F. 2011, in IAU Symposium, Vol. 273, IAU Symposium, ed. D.~{Prasad
  Choudhary} \& K.~G. {Strassmeier}, 89--95

\bibitem[{{Lanza} {et~al.}(2009){Lanza}, {Pagano}, {Leto}, {Messina},
  {Aigrain}, {Alonso}, {Auvergne}, {Baglin}, {Barge}, {Bonomo}, {Boumier},
  {Collier Cameron}, {Comparato}, {Cutispoto}, {de Medeiros}, {Foing},
  {Kaiser}, {Moutou}, {Parihar}, {Silva-Valio}, \&
  {Weiss}}]{2009A&A...493..193L}
{Lanza}, A.~F., {Pagano}, I., {Leto}, G., {et~al.} 2009, \aap, 493, 193

\bibitem[{{Lehtinen} {et~al.}(2011){Lehtinen}, {Jetsu}, {Hackman}, {Kajatkari},
  \& {Henry}}]{2011A&A...527A.136L}
{Lehtinen}, J., {Jetsu}, L., {Hackman}, T., {Kajatkari}, P., \& {Henry}, G.~W.
  2011, \aap, 527, A136

\bibitem[{{Lister} {et~al.}(1999){Lister}, {Collier Cameron}, \&
  {Bartus}}]{1999MNRAS.307..685L}
{Lister}, T.~A., {Collier Cameron}, A., \& {Bartus}, J. 1999, \mnras, 307, 685

\bibitem[{{Maceroni} {et~al.}(1991){Maceroni}, {van't Veer}, \&
  {Vilhu}}]{1991Msngr..66...47M}
{Maceroni}, C., {van't Veer}, F., \& {Vilhu}, O. 1991, The Messenger, 66, 47

\bibitem[{{Maceroni} {et~al.}(1994){Maceroni}, {Vilhu}, {van't Veer}, \& {van
  Hamme}}]{1994A&A...288..529M}
{Maceroni}, C., {Vilhu}, O., {van't Veer}, F., \& {van Hamme}, W. 1994, \aap,
  288, 529

\bibitem[{{Mamajek} \& {Bell}(2014)}]{2014MNRAS.445.2169M}
{Mamajek}, E.~E. \& {Bell}, C.~P.~M. 2014, \mnras, 445, 2169

\bibitem[{{Marsden} {et~al.}(2011{\natexlab{a}}){Marsden}, {Jardine},
  {Ram{\'{\i}}rez V{\'e}lez}, {Alecian}, {Brown}, {Carter}, {Donati},
  {Dunstone}, {Hart}, {Semel}, \& {Waite}}]{2011MNRAS.413.1922M}
{Marsden}, S.~C., {Jardine}, M.~M., {Ram{\'{\i}}rez V{\'e}lez}, J.~C., {et~al.}
  2011{\natexlab{a}}, \mnras, 413, 1922

\bibitem[{{Marsden} {et~al.}(2011{\natexlab{b}}){Marsden}, {Jardine},
  {Ram{\'{\i}}rez V{\'e}lez}, {Alecian}, {Brown}, {Carter}, {Donati},
  {Dunstone}, {Hart}, {Semel}, \& {Waite}}]{2011MNRAS.413.1939M}
{Marsden}, S.~C., {Jardine}, M.~M., {Ram{\'{\i}}rez V{\'e}lez}, J.~C., {et~al.}
  2011{\natexlab{b}}, \mnras, 413, 1939

\bibitem[{{Marsden} {et~al.}(2006){Marsden}, {Mengel}, {Donati}, {Carter},
  {Semel}, \& {Petit}}]{2006ASPC..358..401M}
{Marsden}, S.~C., {Mengel}, M.~W., {Donati}, F., {et~al.} 2006, in Astronomical
  Society of the Pacific Conference Series, Vol. 358, Astronomical Society of
  the Pacific Conference Series, ed. R.~{Casini} \& B.~W. {Lites}, 401

\bibitem[{{Marsden} {et~al.}(2004){Marsden}, {Waite}, {Carter}, \&
  {Donati}}]{2004AN....325..246M}
{Marsden}, S.~C., {Waite}, I.~A., {Carter}, B.~D., \& {Donati}, J.-F. 2004,
  Astronomische Nachrichten, 325, 246

\bibitem[{{Marsden} {et~al.}(2005){Marsden}, {Waite}, {Carter}, \&
  {Donati}}]{2005MNRAS.359..711M}
{Marsden}, S.~C., {Waite}, I.~A., {Carter}, B.~D., \& {Donati}, J.-F. 2005,
  \mnras, 359, 711

\bibitem[{{Meibom} {et~al.}(2011){Meibom}, {Mathieu}, {Stassun}, {Liebesny}, \&
  {Saar}}]{2011ApJ...733..115M}
{Meibom}, S., {Mathieu}, R.~D., {Stassun}, K.~G., {Liebesny}, P., \& {Saar},
  S.~H. 2011, \apj, 733, 115

\bibitem[{{Mentuch} {et~al.}(2008){Mentuch}, {Brandeker}, {van Kerkwijk},
  {Jayawardhana}, \& {Hauschildt}}]{2008ApJ...689.1127M}
{Mentuch}, E., {Brandeker}, A., {van Kerkwijk}, M.~H., {Jayawardhana}, R., \&
  {Hauschildt}, P.~H. 2008, \apj, 689, 1127

\bibitem[{{Moss} {et~al.}(1995){Moss}, {Barker}, {Brandenburg}, \&
  {Tuominen}}]{1995A&A...294..155M}
{Moss}, D., {Barker}, D.~M., {Brandenburg}, A., \& {Tuominen}, I. 1995, \aap,
  294, 155

\bibitem[{{Mosser} {et~al.}(2009){Mosser}, {Baudin}, {Lanza}, {Hulot},
  {Catala}, {Baglin}, \& {Auvergne}}]{2009A&A...506..245M}
{Mosser}, B., {Baudin}, F., {Lanza}, A.~F., {et~al.} 2009, \aap, 506, 245

\bibitem[{{Pallavicini} {et~al.}(1993){Pallavicini}, {Cutispoto}, {Randich}, \&
  {Gratton}}]{1993A&A...267..145P}
{Pallavicini}, R., {Cutispoto}, G., {Randich}, S., \& {Gratton}, R. 1993, \aap,
  267, 145

\bibitem[{{Passos} {et~al.}(2014){Passos}, {Nandy}, {Hazra}, \&
  {Lopes}}]{2014A&A...563A..18P}
{Passos}, D., {Nandy}, D., {Hazra}, S., \& {Lopes}, I. 2014, \aap, 563, A18

\bibitem[{{Paxton} {et~al.}(2011){Paxton}, {Bildsten}, {Dotter}, {Herwig},
  {Lesaffre}, \& {Timmes}}]{2011ApJS..192....3P}
{Paxton}, B., {Bildsten}, L., {Dotter}, A., {et~al.} 2011, \apjs, 192, 3

\bibitem[{{Piluso} {et~al.}(2008){Piluso}, {Lanza}, {Pagano}, {Lanzafame}, \&
  {Donati}}]{2008MNRAS.387..237P}
{Piluso}, N., {Lanza}, A.~F., {Pagano}, I., {Lanzafame}, A.~C., \& {Donati},
  J.-F. 2008, \mnras, 387, 237

\bibitem[{{Piskunov} {et~al.}(1995){Piskunov}, {Kupka}, {Ryabchikova}, {Weiss},
  \& {Jeffery}}]{1995A&AS..112..525P}
{Piskunov}, N.~E., {Kupka}, F., {Ryabchikova}, T.~A., {Weiss}, W.~W., \&
  {Jeffery}, C.~S. 1995, \aaps, 112, 525

\bibitem[{{Press} {et~al.}(1992){Press}, {Teukolsky}, {Vetterling}, \&
  {Flannery}}]{1992nrfa.book.....P}
{Press}, W.~H., {Teukolsky}, S.~A., {Vetterling}, W.~T., \& {Flannery}, B.~P.
  1992, {Numerical recipes in FORTRAN. The art of scientific computing}

\bibitem[{{Reinhold} {et~al.}(2013){Reinhold}, {Reiners}, \&
  {Basri}}]{2013A&A...560A...4R}
{Reinhold}, T., {Reiners}, A., \& {Basri}, G. 2013, \aap, 560, A4

\bibitem[{{Ruediger}(1989)}]{1989drsc.book.....R}
{Ruediger}, G. 1989, {Differential rotation and stellar convection. Sun and the
  solar stars}

\bibitem[{{Semel}(1989)}]{1989A&A...225..456S}
{Semel}, M. 1989, \aap, 225, 456

\bibitem[{{Semel} {et~al.}(1993){Semel}, {Donati}, \&
  {Rees}}]{1993A&A...278..231S}
{Semel}, M., {Donati}, J.-F., \& {Rees}, D.~E. 1993, \aap, 278, 231

\bibitem[{{Silva-Valio}(2008)}]{2008ApJ...683L.179S}
{Silva-Valio}, A. 2008, \apjl, 683, L179

\bibitem[{{Silva-Valio} {et~al.}(2010){Silva-Valio}, {Lanza}, {Alonso}, \&
  {Barge}}]{2010A&A...510A..25S}
{Silva-Valio}, A., {Lanza}, A.~F., {Alonso}, R., \& {Barge}, P. 2010, \aap,
  510, A25

\bibitem[{{Skelly} {et~al.}(2010){Skelly}, {Donati}, {Bouvier}, {Grankin},
  {Unruh}, {Artemenko}, \& {Petrov}}]{2010MNRAS.403..159S}
{Skelly}, M.~B., {Donati}, J.-F., {Bouvier}, J., {et~al.} 2010, \mnras, 403,
  159

\bibitem[{{Solanki}(2003)}]{2003A&ARv..11..153S}
{Solanki}, S.~K. 2003, \aapr, 11, 153

\bibitem[{{Stokes}(1972)}]{1972MNRAS.159..165S}
{Stokes}, N.~R. 1972, \mnras, 159, 165

\bibitem[{{Stout-Batalha} \& {Vogt}(1999)}]{1999ApJS..123..251S}
{Stout-Batalha}, N.~M. \& {Vogt}, S.~S. 1999, \apjs, 123, 251

\bibitem[{{Strassmeier}(2009)}]{2009A&ARv..17..251S}
{Strassmeier}, K.~G. 2009, \aapr, 17, 251

\bibitem[{{Strassmeier} {et~al.}(1997{\natexlab{a}}){Strassmeier}, {Bartus},
  {Cutispoto}, \& {Rodono}}]{1997A&AS..125...11S}
{Strassmeier}, K.~G., {Bartus}, J., {Cutispoto}, G., \& {Rodono}, M.
  1997{\natexlab{a}}, \aaps, 125, 11

\bibitem[{{Strassmeier} {et~al.}(1997{\natexlab{b}}){Strassmeier}, {Boyd},
  {Epand}, \& {Granzer}}]{1997PASP..109..697S}
{Strassmeier}, K.~G., {Boyd}, L.~J., {Epand}, D.~H., \& {Granzer}, T.
  1997{\natexlab{b}}, \pasp, 109, 697

\bibitem[{{Strassmeier} {et~al.}(2013){Strassmeier}, {Carroll}, {Ilyin}, \&
  {J{\"a}rvinen}}]{2013IAUS..294..447S}
{Strassmeier}, K.~G., {Carroll}, T.~A., {Ilyin}, I., \& {J{\"a}rvinen}, S.
  2013, in IAU Symposium, Vol. 294, IAU Symposium, ed. A.~G. {Kosovichev},
  E.~{de Gouveia Dal Pino}, \& Y.~{Yan}, 447--458

\bibitem[{{Strassmeier} {et~al.}(2004){Strassmeier}, {Granzer}, {Weber},
  {Woche}, {Andersen}, {Bartus}, {Bauer}, {Dionies}, {Popow}, {Fechner},
  {Hildebrandt}, {Washuettl}, {Ritter}, {Schwope}, {Staude}, {Paschke},
  {Stolz}, {Serre-Ricart}, {de la Rosa}, \& {Arnay}}]{2004AN....325..527S}
{Strassmeier}, K.~G., {Granzer}, T., {Weber}, M., {et~al.} 2004, Astronomische
  Nachrichten, 325, 527

\bibitem[{{Strassmeier} \& {Rice}(2003)}]{2003A&A...399..315S}
{Strassmeier}, K.~G. \& {Rice}, J.~B. 2003, \aap, 399, 315

\bibitem[{{Tagliaferri} {et~al.}(1994){Tagliaferri}, {Cutispoto},
  {Pallavicini}, {Randich}, \& {Pasquini}}]{1994A&A...285..272T}
{Tagliaferri}, G., {Cutispoto}, G., {Pallavicini}, R., {Randich}, S., \&
  {Pasquini}, L. 1994, \aap, 285, 272

\bibitem[{{Unruh} \& {Collier Cameron}(1997)}]{1997MNRAS.290L..37U}
{Unruh}, Y.~C. \& {Collier Cameron}, A. 1997, \mnras, 290, L37

\bibitem[{{Unruh} {et~al.}(1995){Unruh}, {Collier Cameron}, \&
  {Cutispoto}}]{1995MNRAS.277.1145U}
{Unruh}, Y.~C., {Collier Cameron}, A., \& {Cutispoto}, G. 1995, \mnras, 277,
  1145

\bibitem[{{Warnecke} {et~al.}(2011){Warnecke}, {Brandenburg}, \&
  {Mitra}}]{2011A&A...534A..11W}
{Warnecke}, J., {Brandenburg}, A., \& {Mitra}, D. 2011, \aap, 534, A11

\bibitem[{{Weise} {et~al.}(2010){Weise}, {Launhardt}, {Setiawan}, \&
  {Henning}}]{2010A&A...517A..88W}
{Weise}, P., {Launhardt}, R., {Setiawan}, J., \& {Henning}, T. 2010, \aap, 517,
  A88

\bibitem[{{White} {et~al.}(2007){White}, {Gabor}, \&
  {Hillenbrand}}]{2007AJ....133.2524W}
{White}, R.~J., {Gabor}, J.~M., \& {Hillenbrand}, L.~A. 2007, \aj, 133, 2524

\bibitem[{{Wichmann} {et~al.}(2003){Wichmann}, {Schmitt}, \&
  {Hubrig}}]{2003A&A...399..983W}
{Wichmann}, R., {Schmitt}, J.~H.~M.~M., \& {Hubrig}, S. 2003, \aap, 399, 983

\end{thebibliography}

\end{document}